%% file: rays_mnras_review2_1.tex
\documentclass[useAMS,usenatbib]{mn2e}

\usepackage{graphicx}
\usepackage{natbib}

\usepackage{amsmath}
\usepackage{empheq}
\usepackage{amsfonts}

\usepackage{times}

\usepackage{aas_macros}

\newcommand{\btd}{\bigtriangledown}

\newcommand{\const}{{\rm const}}
\newcommand{\ie}{i.e.\ }
\newcommand{\be}{\begin{equation}}
\newcommand{\ee}{\end{equation}}
\newcommand{\beq}{\begin{eqnarray}}
\newcommand{\eeq}{\end{eqnarray}}

	\newcommand{\beqa}{\begin{eqnarray} \begin{cases}}
	\newcommand{\eeqa}{\end{cases} \end{eqnarray}}

\numberwithin{equation}{section}

	\newtheorem{proposition}{Proposition}

\newcommand{\ba}{\begin{align}}
\newcommand{\ea}{\end{align}}

\newcommand{\vect}[1]{{\bf #1}}

\newcommand{\dd}[2]{\frac{\partial #1}{\partial #2}}

\newcommand{\ddil}[2]{\partial #1 / \partial #2}

\newcommand{\ddp}[3]{
\frac{\partial^{#1} #2}{\partial #3^{#1}}
}

\newcommand{\dder}[2]{
\frac{d #1}{d #2}
}

\newsavebox{\astrutbox}
\sbox{\astrutbox}{\rule[-5pt]{0pt}{20pt}}

\newcommand\eg{e.g.\ }

\newtheorem{corollary}{Corollary}

\title[Acoustic waves generated by point source]{Geometric properties of acoustic waves generated by a point source in solar-like interior: effects of acoustic cut-off frequency.}

\author[S. Zharkov]{S. Zharkov$^{1,2}$\thanks{E-mail: s.zharkov at hull.ac.uk}
\\$^1$Department of Physics and Mathematics, University of Hull, Cottingham Road, Kingston-upon-Hull, HU6~7RX, UK
\\$^2$Mullard Space Science Laboratory,
          University College London, Holmbury St. Mary, Dorking, RH5 6NT,  UK}
\begin{document}

\date{}

\pagerange{\pageref{firstpage}--\pageref{lastpage}} \pubyear{2002}

\maketitle

\label{firstpage}

\begin{abstract}
Acoustic waves generated by a point source in stratified plasma are considered in this paper. Analytical parametric solution for monochromatic source is derived for plane-parallel polytrope model of the solar interior. The solution is used to gain insight into the properties of the generated wavefront as a function of excitation frequency and depth. A slowly varying pressure perturbation moving in upper layers of solar photosphere with supersonic speed is also considered. It is shown to excite acoustic waves putting certain restrictions upon their geometry of the generated wavefront. The results are discussed in relation to flare generated sunquakes. 
\end{abstract}


\section{Introduction}

Ray theory, as form of geometrical optics, has played a crucial role in helioseismology of acoustic oscillations, laying foundations to such methods as time-distance helioseismology \citep{Duvall93, KosDuval1997, Giles2000, DSilva1995, DSilva1996, ThompsonZharkov2008}, acoustic holography \citep{LB2000, LB2004} and far-side imaging \citep{LDfarside}. It has provided us with rich insight into the nature of acoustic wave propagation in the Sun from properties of low-degree modes and Duvall's law used in global helioseismology to averaging setups and focusing information relied upon in local helioseismic techniques. While more accurate numerical methods have since been deployed for computing helioseismic sensitivity kernels, which are used to deduce subsurface properties of the solar interior from observations, these have generally confirmed and built on earlier results from ray-theory \citep{Couvidat2004,Couvidat2006}.

Nonetheless, the use of ray-theory in solar physics has mostly centred on studies of individual rays, or modes, with little attention paid to fully transcribing the initial conditions responsible for generating the family of rays that fully describes the wavefield. For example,  \citet{Bogdan97} has compared the relationship between the modal and time-distance formulations of local helioseismology, but by own admission neglecting to take into account the initial value problem in terms of ray theory. One of the simplest of such cases is a spherical monochromatic point source, which is considered in this paper. 

When applied to standard wave-equation the mathematical method of geometric optics is known by other names such as Debye ray-theory, geometric acoustics, geometric seismics, WKB or semiclassical approximation leading to a fair amount of confusion. In solar physics some of the more rigorous applications of the method can be found in works by \citet[and references therein]{gough93,gough2007}.

Here I use purely mathematical approach as set out by \citet{KravtsovOrlov} for standard wave-equation and generalised for general asymptotic differential operators by \cite{GeoAsymptotics} to explicitly derive and interpret the eikonal and transport equations for non-linear wave equation in the presence of acoustic cut-off frequency.  Finding the full solution of the problem depends on reconstruction of the phase function throughout the whole domain which is accomplished by solving the eikonal equation. The phase function defines the geometry of the problem and can be found when proper initial conditions are prescribed. I then find such solution for the point source using basic polytrope model of the solar interior and consider its basic properties such as wavefront geometry, caustics and surface manifestations. \footnote{Note that there are also other independent means of deriving the eikonal equation such as Fourier transform and stationary phase method \citep{Chapman99,FM}.}

{For physical applications, the results are compared with known acoustic sources in the Sun.} Stochastic turbulent convection in the upper layers of the convective zone is generally thought to generate most of the solar acoustic spectrum. However, another known source of sound waves in the Sun are acoustically active solar flares. In such flares, the energy release in the corona generates observable photospheric ripples that accelerate radially  outward from a source region. This recently discovered phenomena known as 'sunquake'  \citep{kz1998} provides us with a localised example of wave excitation in the Sun. The acoustic nature of sunquake photospheric ripples has been well established and localisation of the source position can be deduced with good precision using helioseismic techniques  such as time-distance analysis  \citep{kz1998,K2006,Kosovichev2007} and acoustic holography \cite{Donea1999, DL2005, LD2008}. In particular, the egression computation in acoustic holography uses the Green's function built on the assumption of a monochromatic point source generating downward propagating waves from given depth in the solar atmosphere. 

The paper is organised as follows the mathematical formalism for asymptotical treatment and solution of the solar wave equation, splitting it into eikonal and transport equations is described in Section \ref{s:method}. The link between general solution and initial conditions is addressed in  \ref{sec:inicond}. The solution for eikonal equation for a point source is  derived in Section \ref{sec:polysolve} including the expression for the Jacobian. Possible applications and limitations of the derived solutions to sun-quakes and basic properties of moving source wave-fronts are discussed in Section \ref{sec:discuss}. 

\begin{figure}
\centering
\includegraphics[width=8cm]{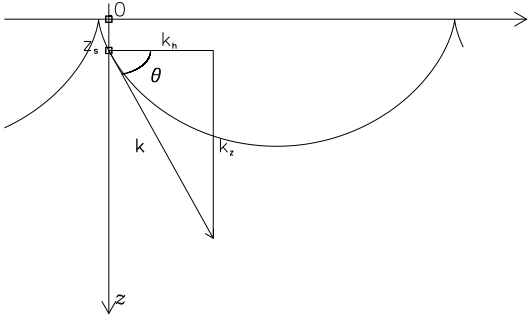}
\caption{Spatial geometry of the problem \label{fig:geo_setup}}
\end{figure}
%
%
%
%
%
%
%
%
%
\input{method}

\input{initial_conditions}
\section{Polytrope} \label{sec:polysolve}
\subsection{Individual ray solution} 
\label{app:ray_solve}
Let us now consider the reduced system in $2+1$ dimensions with Hamiltonian (\ref{eq:Hamiltonian})  independent of horizontal coordinate and time. The characteristic equations (\ref{eq:HJsystem}) are then considerably simplified:
\begin{subequations} \label{eq:acrays}
\begin{empheq}[left=\empheqlbrace]{align}
\frac{dx}{d\tau}=k_h;  \frac{dz}{d\tau} & =  k_z;  \\ 
\frac{dt}{d\tau}=\frac{\omega}{c^2};  
\frac{k_h}{d\tau}=0; \frac{d\omega}{d\tau} & =  0;  
\\ k_h^2+k_z^2 - \frac{\omega^2 -{\omega}_{ac}^2}{c^2} & =  0. 
\end{empheq}
\end{subequations}
When $k_h=0,$ the corresponding ray propagates purely vertically ($x=\const$). It is easy to see that in terms of the reduction described for spherical source in Section  \ref{s:ini_mono_spherical} this ray travels along the rotation axis.

Consider polytrope model with adiabatic sound-speed and acoustic-cutoff frequency depending on depth only:
\beq
c^2(z)=\frac{gz}{m}, \ \omega_{ac}^2(z)=\frac{g(m+2)}{4z}, \label{eq:poly}
\eeq
where $g$ is the gravitational constant (the value of $g=2.67\times 10^{-4} {\rm \ Mm/s^2}$ is used in our calculations), and $m$ is the polytrope index. Let us discuss some general properties of such model and eikonal solving associated rays.

A single ray with horizontal wavenumber $k_h$ and frequency $\omega$ in this model, in general, will have two turning points, $z_u$  and $z_l$ (upper and lower), determined by the condition $k_z=0$. In the exceptional case of ray propagating purely vertically, $k_h=0$, the lower turning point can be viewed as located at infinite depth. Otherwise, using (\ref{eq:poly}) one obtains:
\begin{empheq}{align}
\frac{z^2k_z^2}{k_h^2} & =-z^2+\frac{\omega^2 m}{k_h^2g}z-\frac{m(m+2)}{4k_h^2} \nonumber \\
 & =(z-z_u)(z_l-z)=b^2-(z-a)^2,  
\label{eq:poly_kz_1}
\end{empheq}
where $z_u, z_l$ are the upper and lower turning points and $a(k_h, \omega)=(z_u+z_l)/2, b(k_h, \omega)=(z_l-z_u)/2.$ These can also be written as 
$$a(k_h, \omega)=\frac 1 2 \frac{\omega^2 m}{k_h^2 g}, b(k_h, \omega)=\sqrt{a^2(k_h, \omega)-\frac{m(m+2)}{4k_h^2}}.$$  
Note that the upper and lower turning points coincide when $b=0.$ 


Furthermore, in this model the wave-vector length, $k^2=(\omega^2-\omega_{ac}^2)/c^2,$ viewed as function of $z$ only, has a maximum at $z_E=g(m+2) / 2\omega^2$ determined by $\omega_{ac}^2(z_E)=\omega^2 / 2$. This value is important due to the following more general statement:
\begin{proposition}
With Hamiltonian $H$ in \eqref{eq:Hamiltonian} independent of $x$ and $t$, 
every ray solving (\ref{eq:HJsystem}) that has two distinct turning point depths, $z_u$ and $z_l$, will pass through at least one point with $z=z_E,$ where $z_E$ is such that 
$$\left. \dd{}z\left(\frac{\omega^2-\omega_{ac}^2}{c^2} \right)\right\arrowvert_{z=z_E}=0.$$
\label{thm:prop}
\end{proposition}
Indeed,  according to (\ref{eq:HJsystem}), 
$$\dder{k_z}\tau=-\dd H z= \dd{}z\left(\frac{\omega^2-\omega_{ac}^2}{c^2} \right)$$. 
But this is the rate of change of vertical wavenumber along the ray, and since $k_z(\tau)$ is smooth and becomes zero at turning points, it will have an extremum between them.
\begin{corollary}
\label{thm:corr}
If for a given frequency there exists a unique ''partition depth'' $z=z_E$ such that $$\left. \dd H z \right\arrowvert_{z=z_E}=0,$$ then for all rays with distinct upper and lower turning points the following inequalities are true:
$z_u < z_E, z_l > z_E.$ 
Ray starting at depth $z_E$ with $$k_h^2=\left.\frac{\omega^2-\omega^2_{ac}}{c^2}\right\arrowvert_{z=z_E}$$ will propagate horizontally.
\end{corollary}

In the polytrope model, $b$ becomes zero only when ray is initialised at depth $z_E(\omega)$ with $k_h={\omega^2m}/{2gz_E}$. It follows that under these conditions $a=z_E.$

\begin{figure*}
\centering
\includegraphics[width=7.7cm]{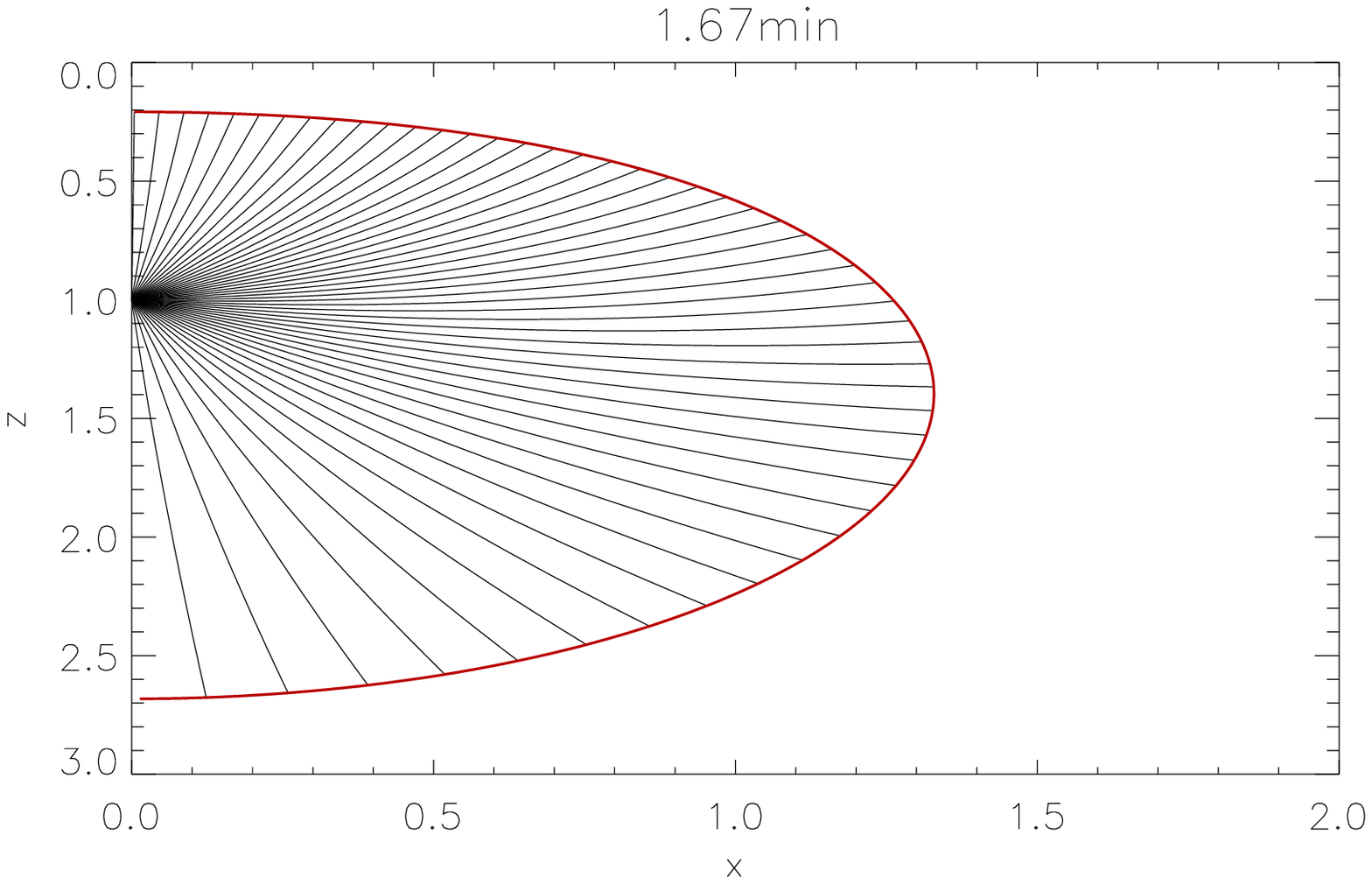} \includegraphics[width=7.7cm]{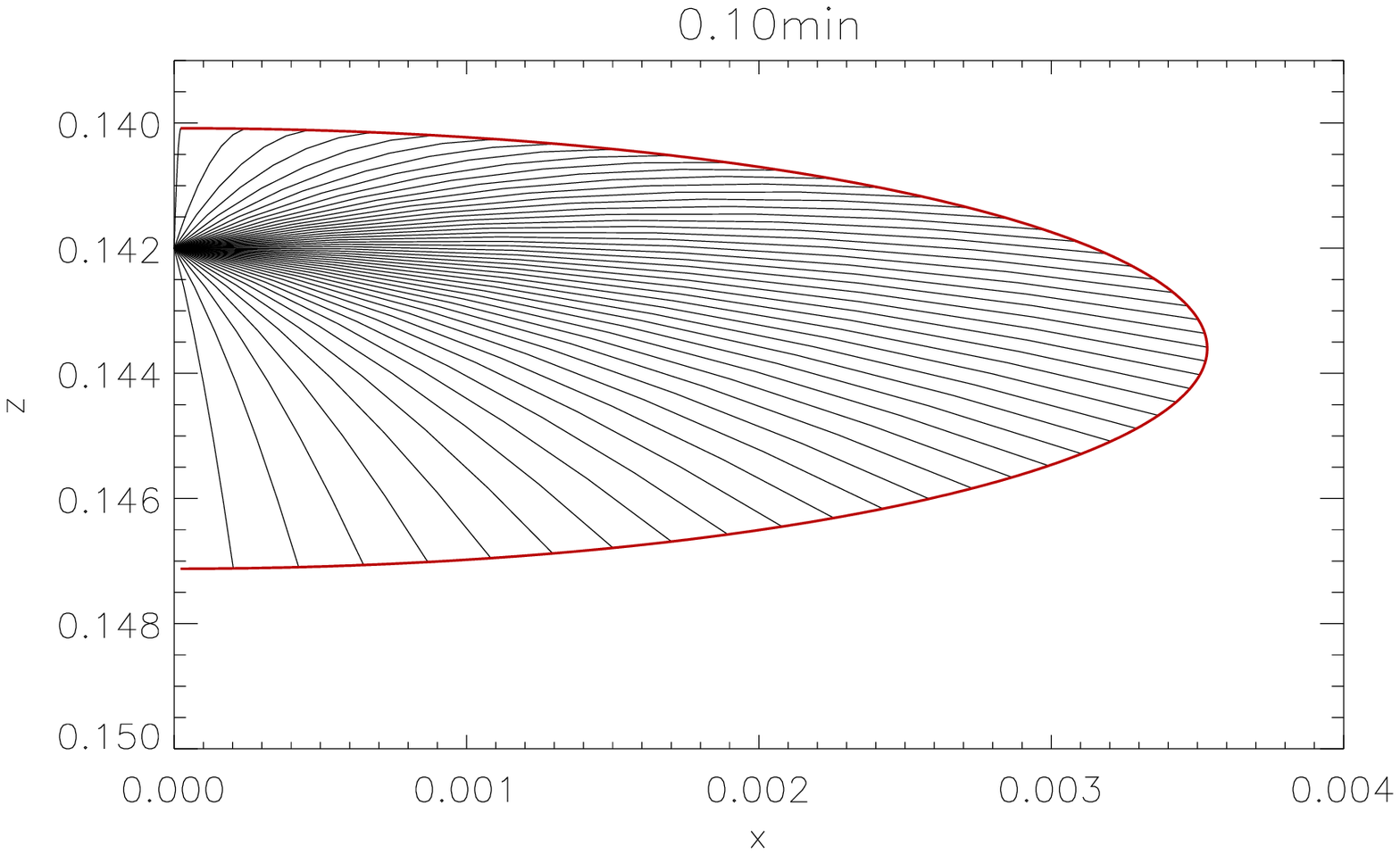}  \\
\includegraphics[width=7.7cm]{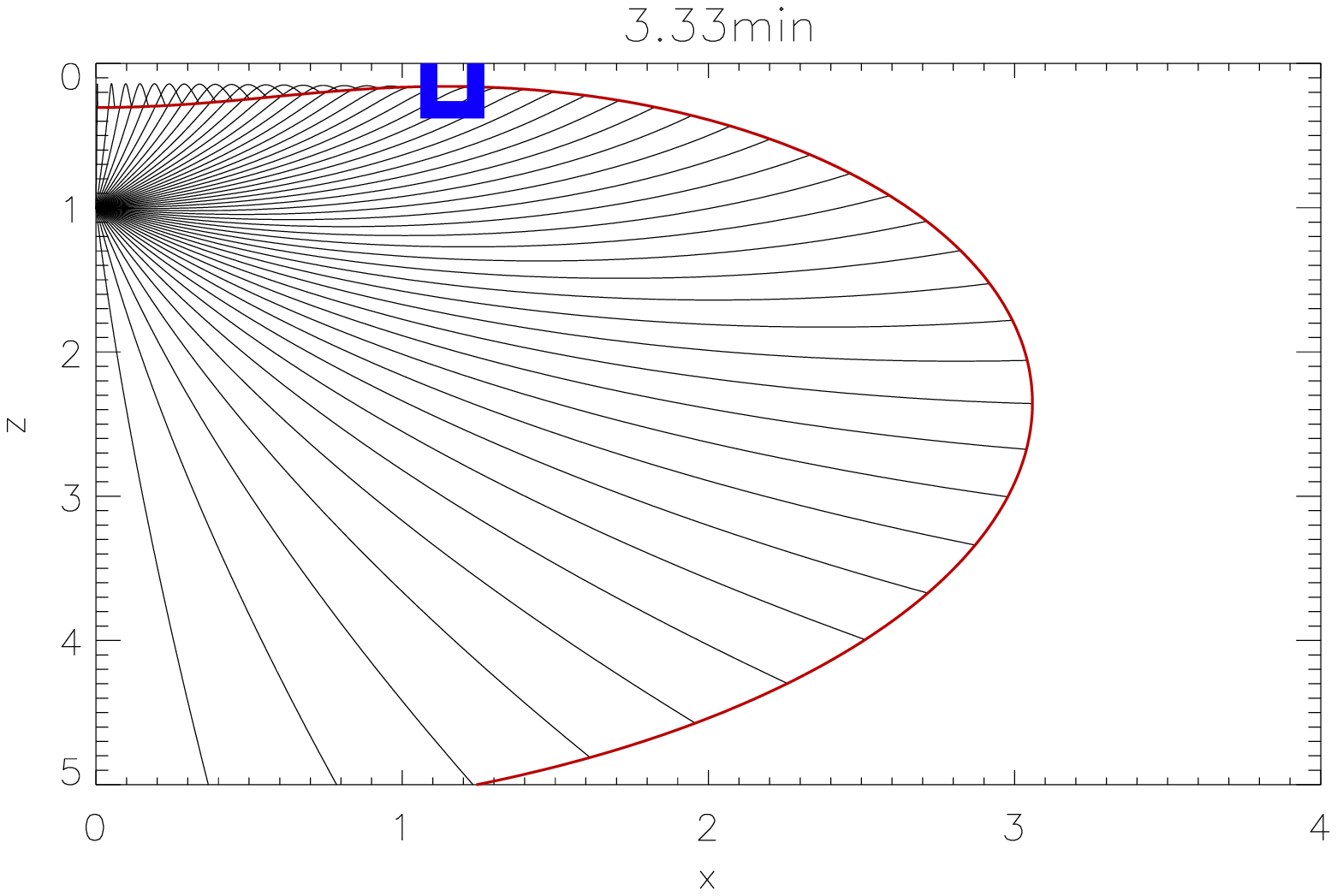} \includegraphics[width=7.7cm]{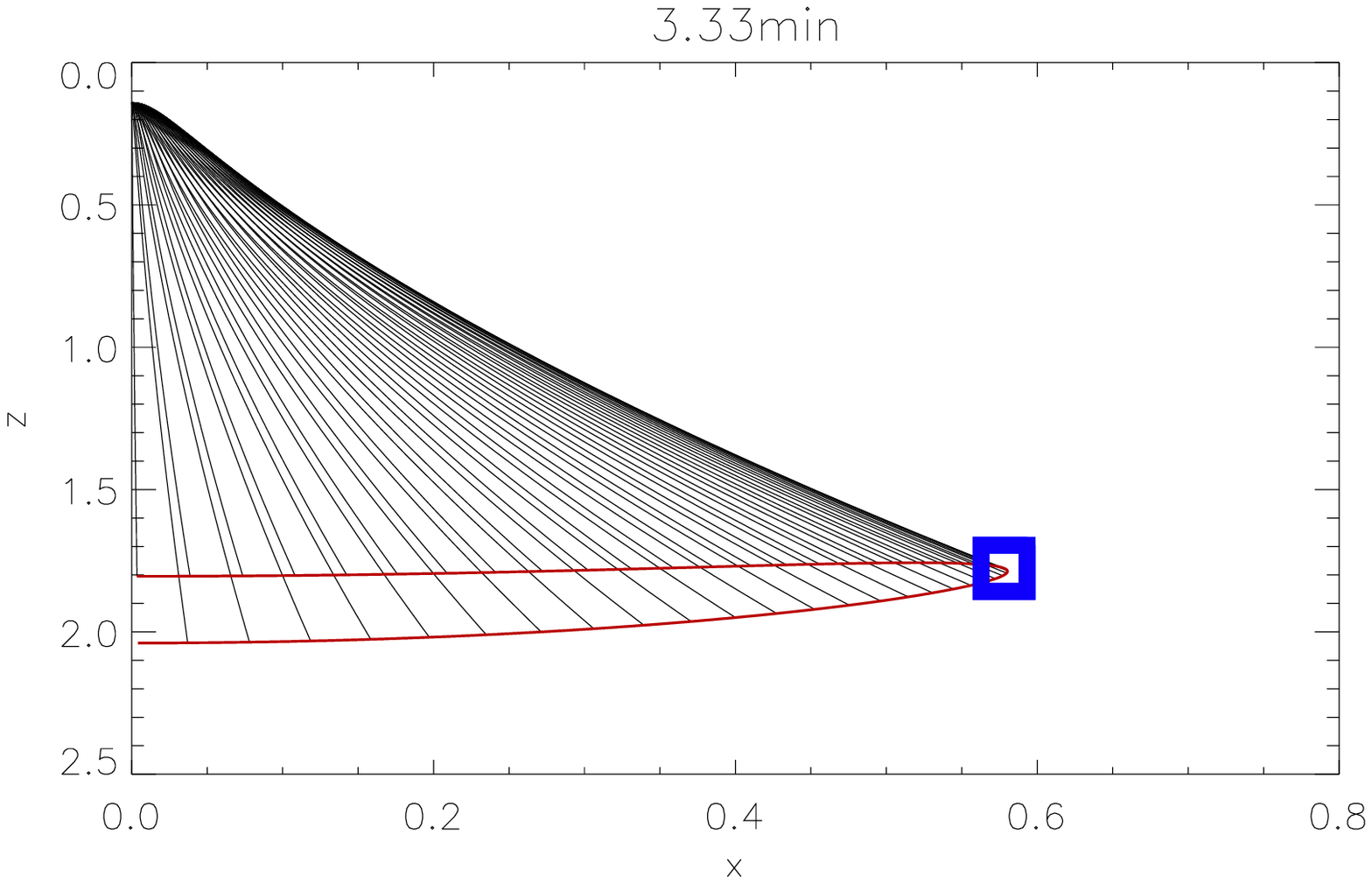} \\
\includegraphics[width=7.7cm]{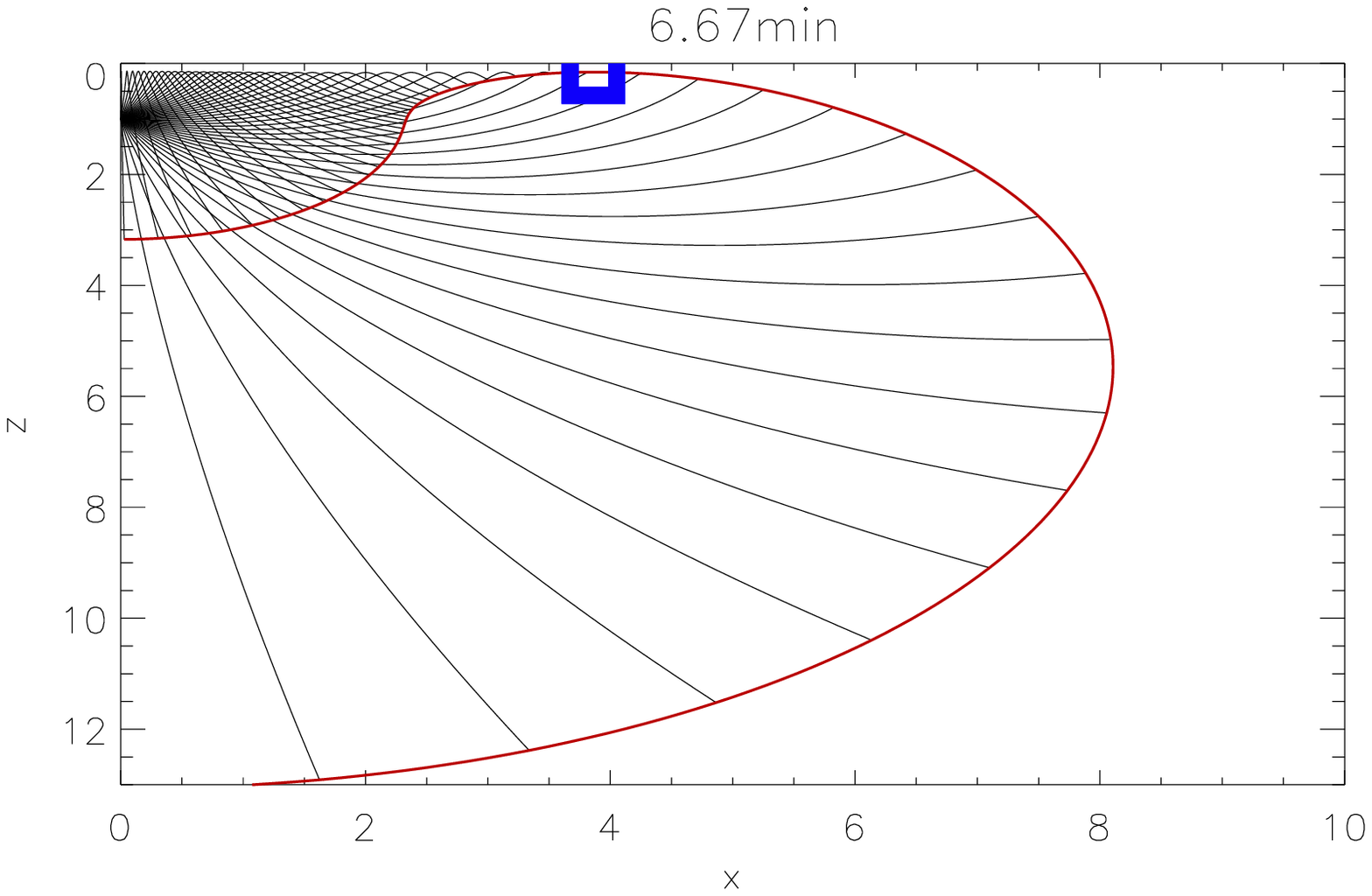} \includegraphics[width=7.7cm]{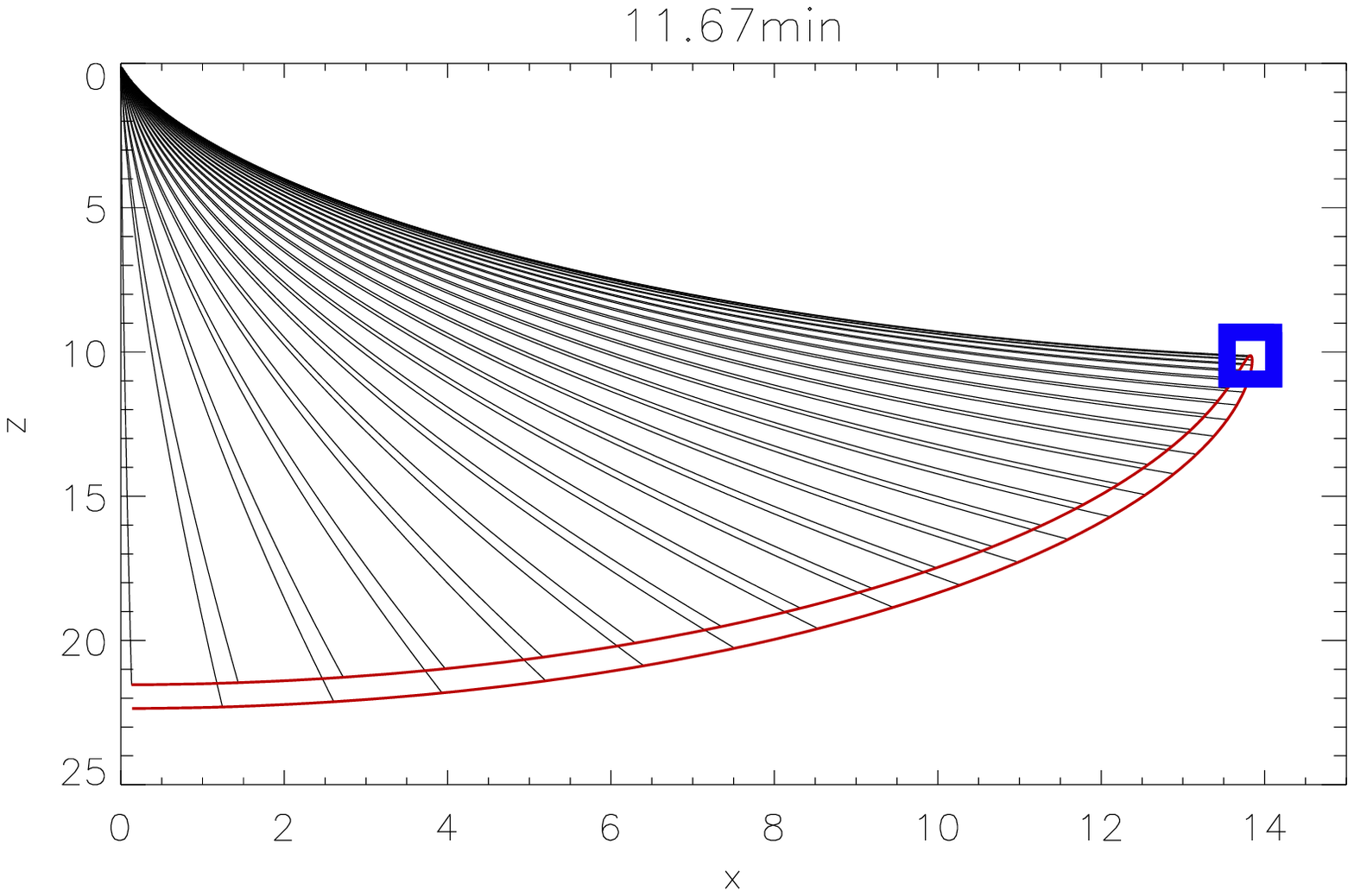} \\
\includegraphics[width=7.7cm]{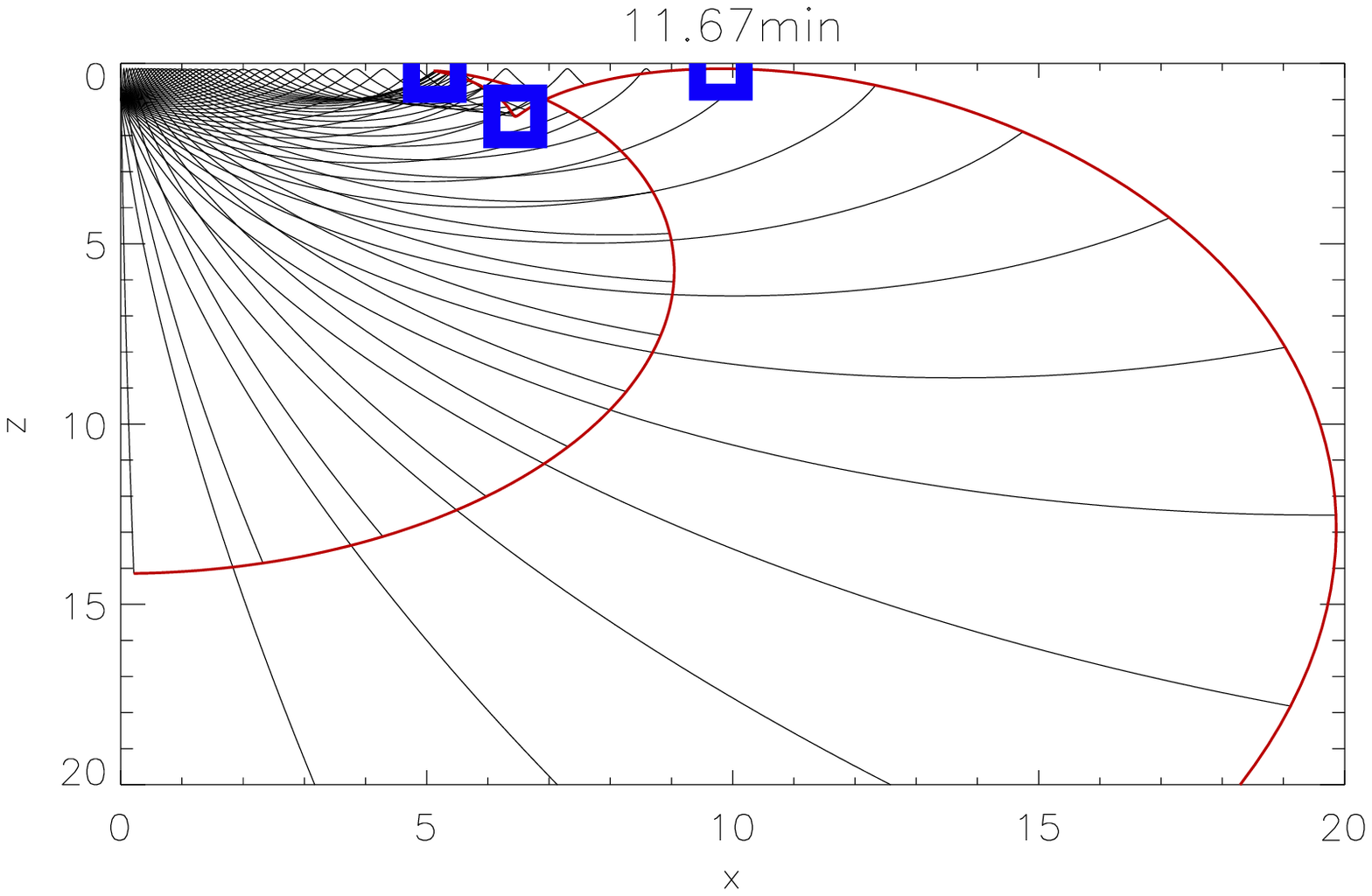} \includegraphics[width=7.7cm]{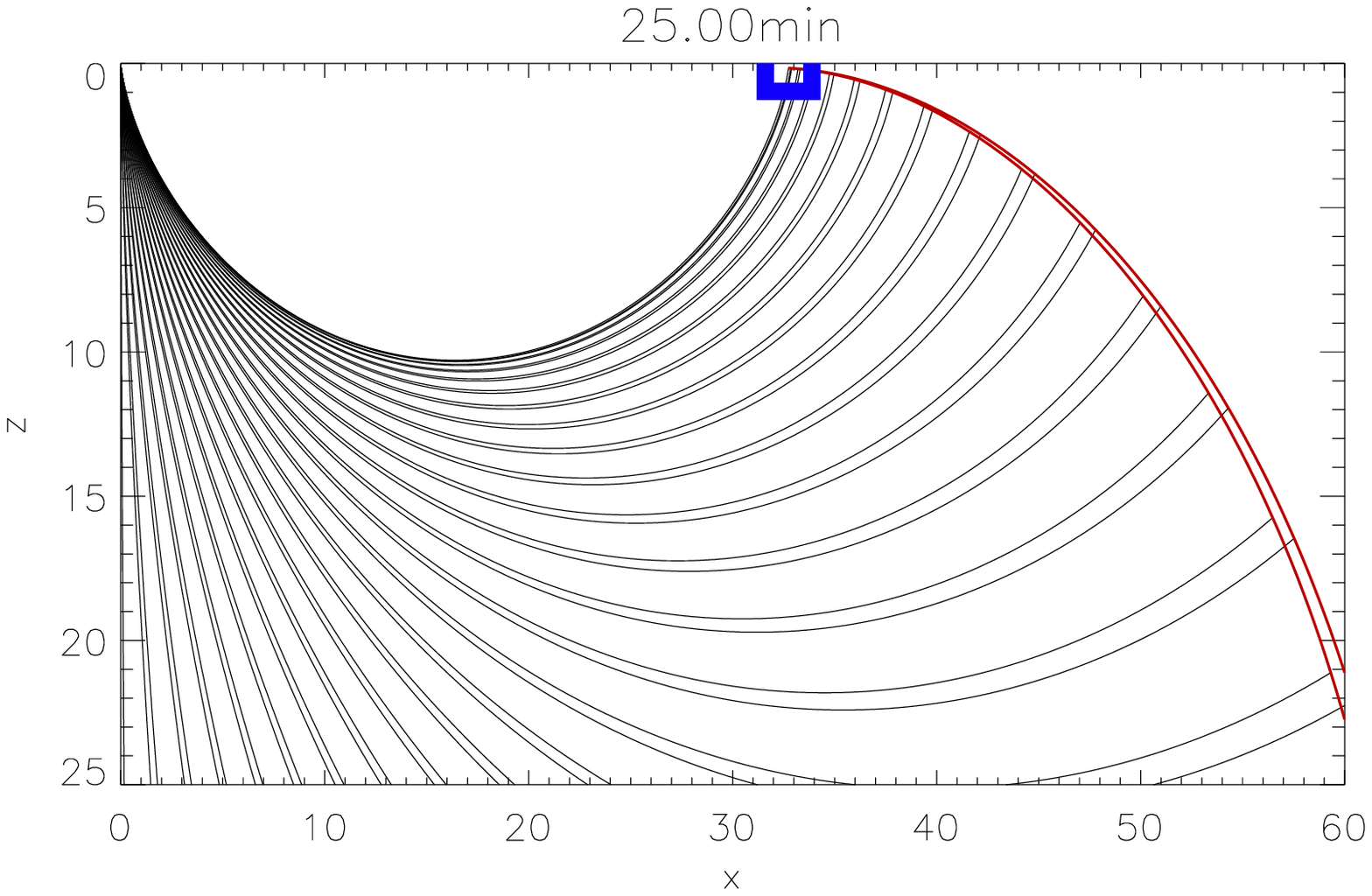}
\caption{Acoustic wavefront at different times generated by monochromatic point source with frequency $\nu=6.5$ mHz with $z_s=1$ Mm {\em (left)} and  $z_s=.142$ Mm  {\em (right)}. Rays start at $t_0=0$, slice times are given in plot titles, geometric wavefront is plotted in red, with selected time-limited rays in black. Caustic points are marked by blue squares. The polytrope index is taken as $m=\frac 3 2.$ The surface is located at $z=0.$ 
Note that the right column corresponds to the case when the source is placed in the region where variability of the model is high, so that the 
eikonal
approximation becomes invalid. Indeed there are reasons to think that such a solution does not represent a physically realistic scenario (C. Lindsey - private communication). See Section \ref{sec:applicability} for discussion of applicability.
\label{fig:wavefront} }
\end{figure*}
%
%

Since $ \omega/{c^2} \neq 0$ in the domain of interest, $z>0$, one can rewrite system (\ref{eq:acrays}) using group travel time, $t_g$, as the parameter along the ray, with $t=t_g+t_0$, to obtain
\begin{subequations} \label{equ:poly_system}
\begin{empheq}[left=\empheqlbrace]{align}
\dder{x}{t_g} & =\frac{gz}{\omega m} k_h, \ \dder {z}{t_g}=\frac{gz}{\omega m} k_z \\
\dder{\omega}{t_g} & =0, \ \dder{k_h}{t_g}=0,  \\
\frac{z^2k_z^2}{k_h^2} & =b^2-(z-a)^2
\end{empheq}
\end{subequations}
Therefore when $b \neq 0,$ 
$$\sin^{-1}\left( \frac{z-a}{b} \right)=\frac{k_h g}{\omega m} t_g + C_z,$$ hence, 
\begin{subequations}
\begin{align}
z & =a+b \sin \alpha, \label{eq_poly_gensolution_z} \\
x & =\frac{1}{2}\frac{\omega}{k_h}t_g-b \cos \alpha +C_x,
\label{eq:poly_gensolution}
\end{align}
\end{subequations}
where $\alpha= t_g {k_h g}/{\omega m} + C_z$ and  integration constants $C_z$ and $C_x$ are determined from the initial conditions. For $b=0$ it is easily checked that $z=z_E=a, x={\omega}t_g/{2k_h}+C_x,$ \ie (\ref{eq:poly_gensolution}) still holds. In the case when $k_h=0$,  $x=\const,$ $z= g t_g^2 /  {4m} \pm C_z t_g \sqrt{g/m}+C_z^2,$ where the choice of sign depends on the direction of the ray propagation and $C_z$ is determined from the initial conditions.

Using group travel time as a parameter along the ray leads to a following interpretation. Let $(\xi, t_0)$ be the coordinates on the initial surface with the field described there as 
$$\Psi_0=A_0^0(\xi, t_0) e^{i\varphi_0(\xi, t_0)},$$ such that $A_0^0=0$ when $t_0 < 0$. 
Then from (\ref{eq:trans0_sol}) it follows that $A_0(\xi, t_0, t_g)=0$ along the ray when $t=t_0+t_g<t_g.$ Hence, group travel time is the time it takes for the initial perturbation to travel to a point along the ray and surfaces $t_g=\const$ represent wavefront snapshots at a particular time \citep{KravtsovOrlov}. 

Using the monochromatic spherical source initial conditions (\ref{eq:ini_sc}-\ref{eq:ini}) with $x_s=0$
the ray system can be  solved in terms of coordinates $(\theta, t_0,  t_g)$ as shown in Appendix \ref{sec:appendix}. Given source depth, $z_s$, the solution is written in terms of take-off angle, $\theta,$ and group travel time $t_g.$ The group travel time is expressed via a parameter $\alpha(\theta, t_g)$ and its initial value $\alpha_0(\theta, z_s).$ Solution for $\alpha_0$ is given in (\ref{eq:alpha_0_sol}), which reveals the importance of the source position relative to the turning point partition depth, $z_E$. Formulas (\ref{eq:alpha_x}-\ref{eq:alpha_alpha0}) together with the system  (\ref{eq:sol_phase_var}) provide the immersion of the phase function graph into phase space: $(\theta, t_0,  t_g)  \hookrightarrow \Omega M.$  Formulas (\ref{eq:alpha_x}-\ref{eq:alpha_alpha}) themselves represent the projection of such map to $M$: $(\theta, t_0,  t_g) \mapsto (x, z, t).$ Due to the simple translational relationship (\ref{eq:alpha_tg}) between the group travel time and time and initial time variables, when considering eikonal solution one is justified in simply fixing $t_0.$  Then for $\theta=\const,$  (\ref{eq:alpha_x}-\ref{eq:alpha_alpha}) provide parametric ray solution as a function of $t_g$, while taking $t_g=\const$  gives geometric wavefront at the time, $t=t_g+t_0,$ as function of take-off angle $\theta.$

Examples of the solutions in terms of ray and wavefront propagation for monochromatic spherical sources placed at two different depths are presented in Figure  \ref{fig:wavefront}. In both cases, the polytrope index is set as $\frac 3 2, t_0=0,$ and source frequency, $\nu= \omega / {2\pi},$ is 6.5 mHz. Each panel represents a spatial snapshot of the eikonal system at a particular time which is given in the panel title. The wavefront is plotted as thick red curve, selected ray paths up to the time indicated are plotted in black. The left column corresponds to the case $z_s= 1{\rm \ Mm} > z_E,$ right column shows snapshots for $z_s= 0.142{\rm \ Mm} < z_E.$ 
After simultaneously leaving the source, rays propagate up ($\theta<0$) and down ($\theta>0$) from the source sweeping out the generated wavefront. Near the surface, at the upper turning points rays reflect back to the interior due to acoustic-cutoff frequency increase. In the interior the rays are reflected back to the surface at lower turning points. As different rays reach turning points at different times, the wavefront gets deformed. This is discussed in the next Section. When source is located below the partition depth $z_E$ (left column), wavefront reaches the surface with the rays going up from the source and then propagates radially outward in the near-surface layers with the second bounce surface manifestation appearing later. When source is located above $z_E,$ the situation is different: after initial surface bounce for rays going up from the source, the reflected wavefront goes below the surface mirroring the behaviour of the rays going down, before reappearing some distance away from the source and then travelling radially outward in the near surface layers. This is further discussed in Section \ref{sec:wave_surface}.


\begin{figure}
\includegraphics[width=8.5cm]{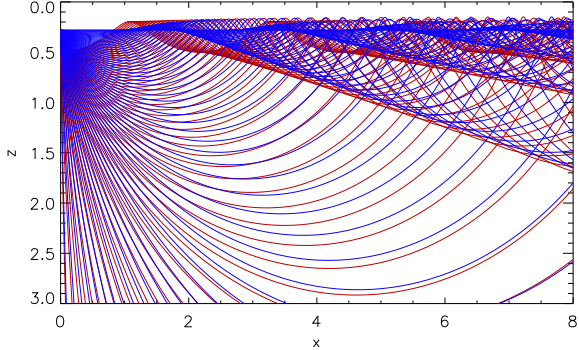}
\caption{Downward propagating rays with $\nu=6.5$ mHz, $m=\frac 3 2$ for point-sources at depths of 1 Mm {\em (blue)} and 2 Mm {\em (red)}. Envelopes of the ray families representing the caustics are clearly seen.
\label{fig:rays}}
\end{figure}
\subsection{Caustics}
Figure \ref{fig:rays} shows two families of rays generated by semi-spherical point sources located at 1 and 2 Mm depths. In both cases, the downward pump scenario\footnote{see the last paragraph of Section \ref{s:ini_mono_spherical}} is represented, i.e. $A_0^0(\theta, t_0)=0$ when $(\theta \geq 0),$ so only downward propagating rays are plotted.
It can be seen that near the upper and the consecutive lower turning points the density of the projected onto $M$ characteristic curves (i.e. rays) increases with ray families becoming enveloped by ''caustics''. Caustics are defined as surfaces where the Jacobian of the coordinate transform from ray to Cartesian coordinates becomes zero. This implies that at such points the mapping is not unique, \ie two or more rays pass through the same point in $(x, z, t)$ space. It can be checked that, under the initial conditions considered here, this is equivalent to the wave-vector being tangent to the wave-front. As wavefront propagates in the direction of it's tangent and rays travel in the direction of the wave-vector, it follows that such points locally envelop the wavefront and ray-paths. Caustics are often, though not always, associated with the singularities of the generated wavefronts such as cusps. See  \citep{KravtsovOrlov,KravstovOrlovCaustics} and references therein for more details.

\begin{figure*}
\centering
\includegraphics[width=6.7cm]{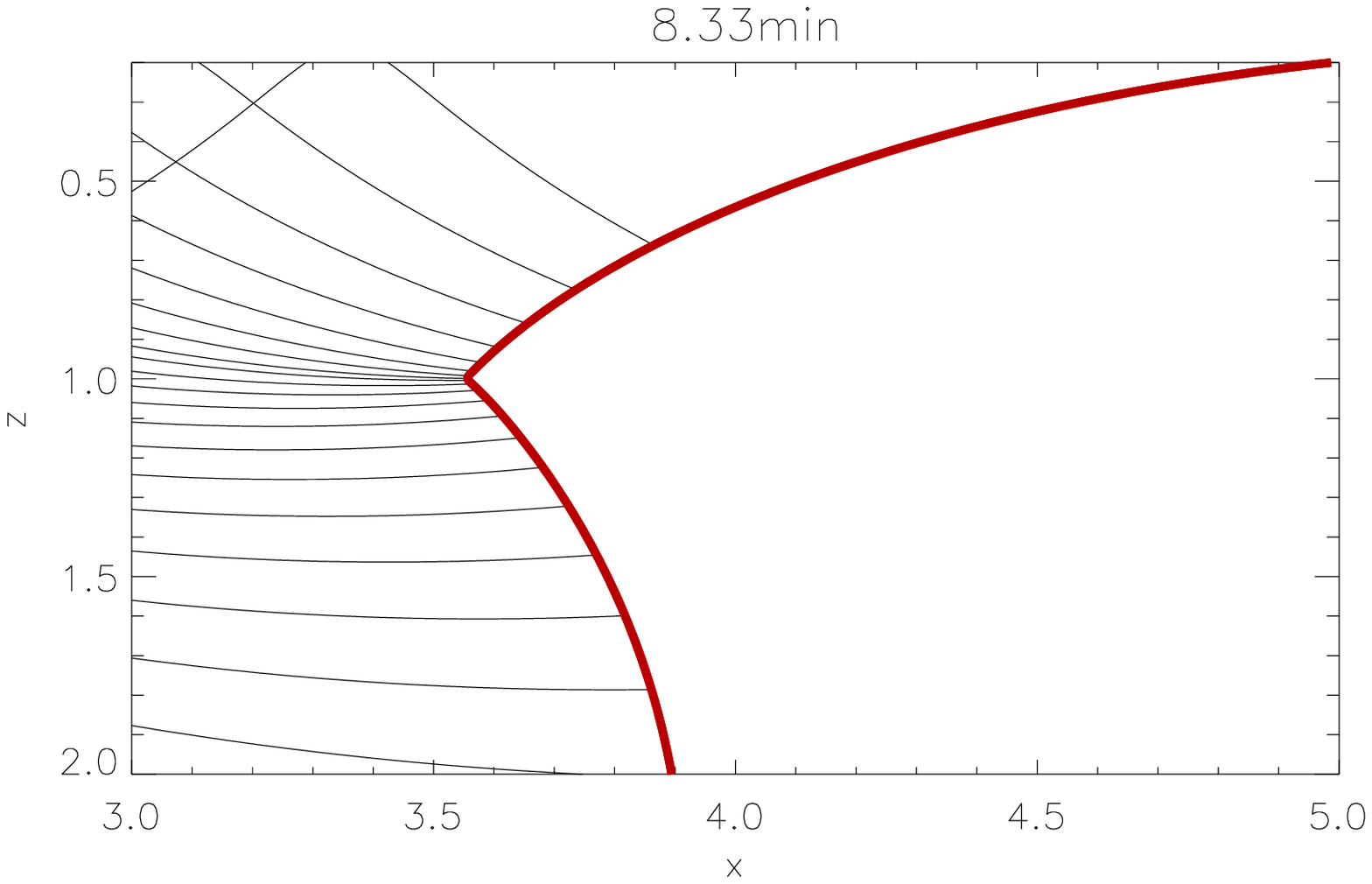} 
\includegraphics[width=6.7cm]{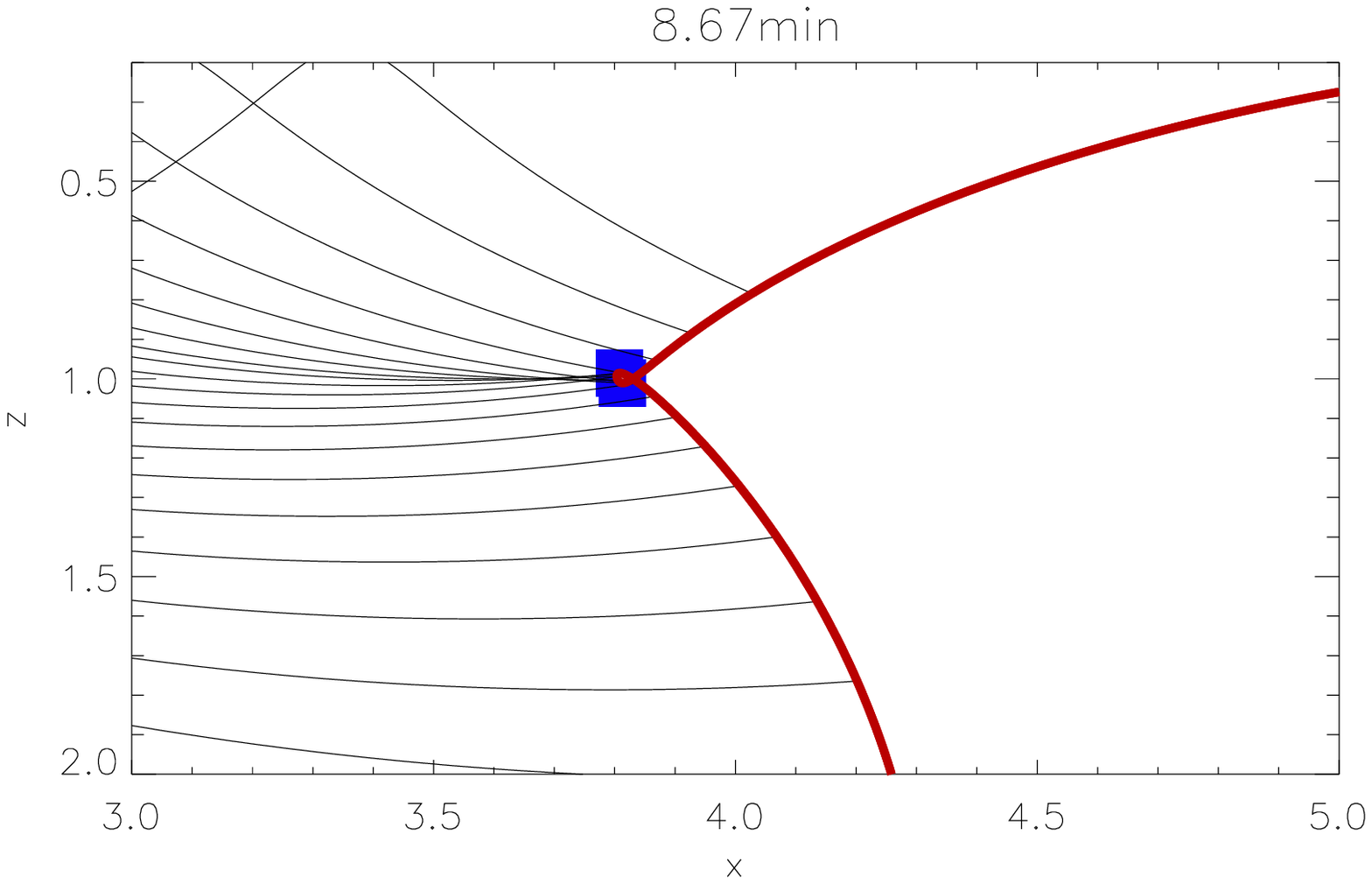} \\
\includegraphics[width=6.7cm]{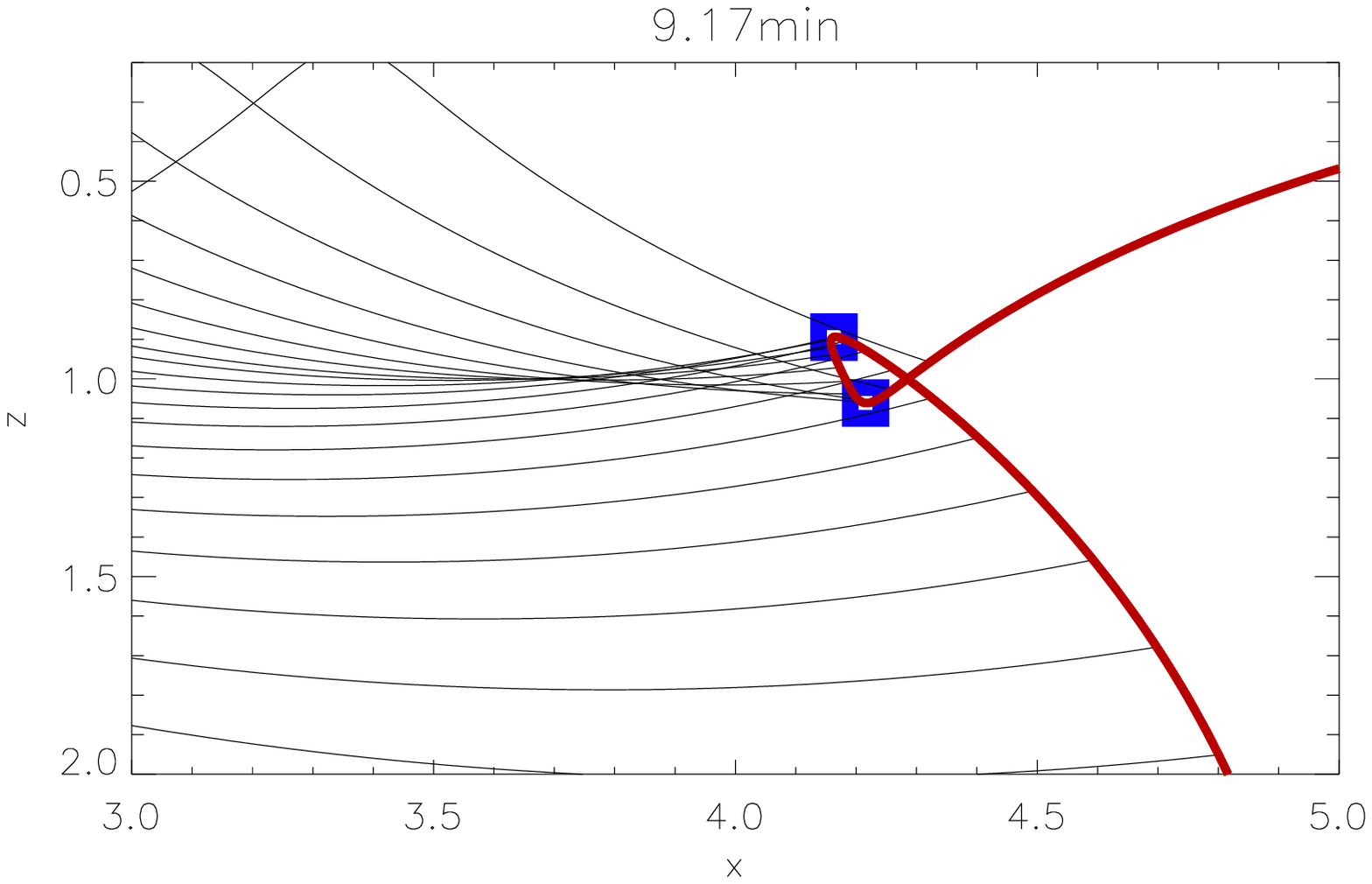} 
\includegraphics[width=6.7cm]{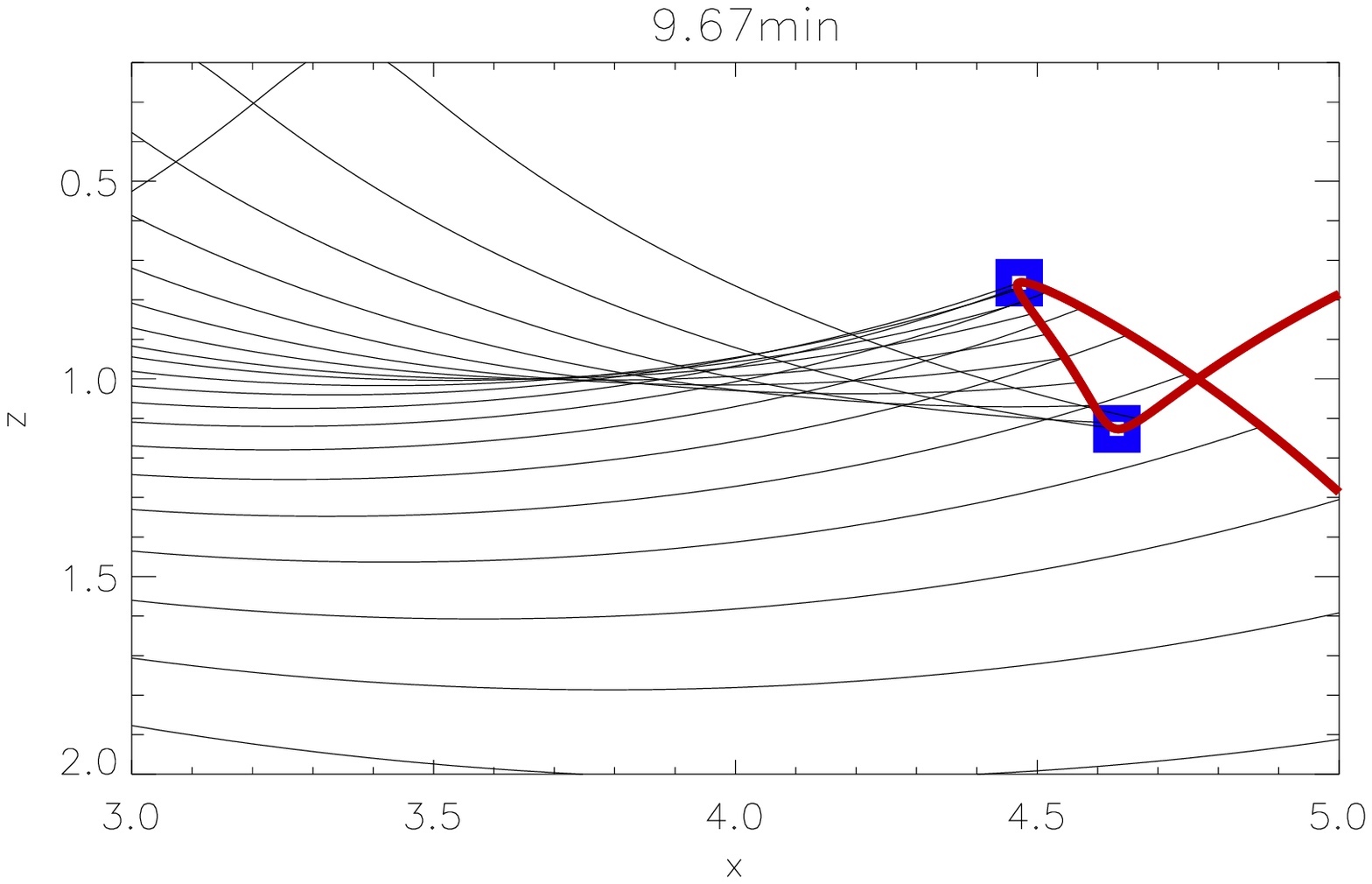}
\caption{Formation of the wavefront singularity and caustics near lower turning points, for $z_s=1$ Mm, $\nu=6.5$ mHz. The caustic is formed when ''up'' rays after being reflected near surface, intersect ''down'' rays travelling up to the surface after passing the lower turning point.
\label{fig:caus_lowtp} }
\end{figure*} 
For $\theta \in \left( - \frac \pi 2, \frac \pi 2 \right),$ analytical expression of the Jacobian of the transform from ray-to-Cartesian coordinates is derived in Appendix \ref{app:jacobian} (see (\ref{eq:jacobiana})).
Evaluating $J(\theta, \alpha)$ at lower turning points, $\alpha= \frac \pi 2 + 2 n \pi,$ gives $J(\theta, \frac \pi 2 + 2 n \pi)=- z_l^3  {\tan\theta} / b$, and at the upper: $J(\theta, -\frac \pi 2 + 2 n \pi)= z_u^3 {\tan\theta} / b$. It follows then that the turning points are on caustics only when $\theta=0.$ Moreover, 
$$J(0, \alpha)=b z_s \cos\alpha\left(1 - \frac \xi 2 \dd{\alpha_0}\theta (0)\right)= - 2b \frac{z_s}{\xi-2} \cos \alpha,$$ 
so that the ray with $\theta=0$ touches caustics only at its turning points. In particular, this means that the initial value of the Jacobian is zero, which according to (\ref{eq:trans0_sol})  indicates that in zero-order approximation the amplitude of the wave-filed will be zero along the ray. When take-off angle is non-zero, the fact that the Jacobian along the ray takes different signs at upper and lower turning points suggests that there is at least one caustic point situated in between.

The caustic points, computed numerically using (\ref{eq:jacobiana}), at given times are plotted as blue squares in Figure  \ref{fig:wavefront}. In both cases caustics are formed after rays went through the turning points. When $z_s<z_E$, right column, the first caustic is formed at the interface of the downward propagating front and the reflected from the surface following front. Once this caustic comes back to the surface (the bottom plot), it then propagates near the surface away from the source.
In the case when $z_s>z_E,$ left column,  the first caustic point is formed near the surface when the ray with $\theta=0$ passes through the upper turning point. The point then travels radially in the near-surface layers away from the source. The bottom image shows the appearance of the next set of caustic points, one of which is again located near the surface and is associated with the second bounce waves at the surface. The others correspond to wavefront cusp formed near lower turning points. A close-up example of the formation of caustics associated with the wave-front cusp is shown in the Figure \ref{fig:caus_lowtp}.

\begin{figure}
\includegraphics[width=8.5cm]{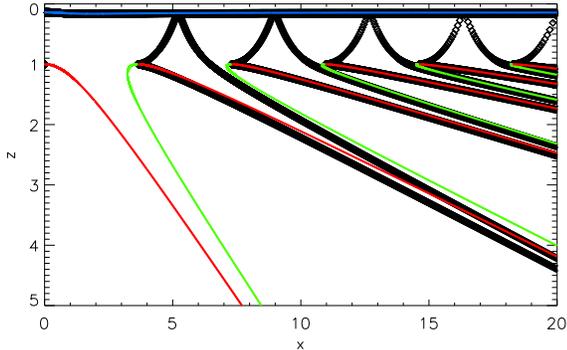}
\caption{Caustics and turning points: projection of caustics onto spatial plane are thick black lines. Lower turning points of rays going down from the source are drawn in red, with lower turning points for rays going up from the source drawn in green. Upper turning points are in purple.
\label{fig:caustics_turningpoints}}
\end{figure}

In Figure \ref{fig:caustics_turningpoints}  locations of the caustics and turning points are compared. It can be seen that the lower caustics are associated with lower turning points for second and further bounces. These caustics are quite close to the locations of the turning point, though not exact except for $\theta=0$. The caustics formed near the surface are very close to rays' upper turning points. Moreover, near surface caustics are generally quite basic, \eg not associated with wavefront singularities, and are formed when only two rays (one after being reflected and another coming up to its reflection point) intersect. 

The results presented in this Section agree well with the analysis of the  numerical simulation of acoustic wave propagation in the solar interior \citep{shelyag2009}. Further analysis is needed to rigorously determine types of the caustics generated by a spherical point source and hence describe acoustic field behaviour in their neighbourhood. 

\subsection{Wavefield at the surface} \label{sec:wave_surface}
As no direct observations of the Sun below the photosphere are available, the surface manifestation would be a primary observable for any acoustic wave-field generated in the interior or near the surface.
Let  $x_{surf}(\theta \arrowvert \omega, z_s)=x_{surf}(\theta)$ denote the horizontal distance from the source to the first appearance of the ray on the surface, \ie its first upper turning point. 
As shown in Appendix \ref{sec:appsolve}, upper turning points correspond to $\alpha=-\frac \pi 2 + 2 \pi n,$ so the first such point away from the source depends on the choice of initial value, $\alpha_0.$ 

When source is placed at or below $z_E,$ $z_s \geq z_E,$ the situation is straight forward with $\alpha_0(\theta) \in \left( -\frac \pi 2, \frac{3\pi}{2} \right)$ for $\theta \in \left( -\frac \pi 2, \frac \pi 2\right),$ so that the first upper turning point along every ray corresponds to $\alpha=\frac 3 2 \pi$. Then
\be
x_{surf}(\theta)
= z_s \frac \xi {2 \cos^2\theta} \left(\frac 3 2 \pi - \alpha_0 \right) +z_s\tan\theta, 
\label{equ:xsurf}
\ee
where 
$$\xi= \left.\frac{\omega^2}{\omega^2-\omega^2_{ac}}\right\arrowvert_{z=z_s}$$ 
(see Appendix \ref{sec:appsolve} for more details). The corresponding group travel time, $t_{g, {\rm surf}},$ the time it takes for the perturbation to travel along the ray from the source to ray's first upper turning point, is given as 
\be t_{g,  {\rm surf}}=\frac{\omega m}{k_h g} \left( \frac 3 2 \pi - \alpha_0 \right). \label{eq:surf_tg}
\ee
When source is placed above $z_E$, \ie $z_s<z_E$, $\alpha_0$ is less or equal to $-\frac \pi 2$ when $\theta$ is negative (rays going up from the source), and greater than or equal to  $-\frac \pi 2$ when $\theta$ is positive. Thus, 
\[
x_{surf}(\theta)=\begin{cases}
      z_s \left( \frac 1 2 \frac \xi {\cos^2\theta} \left(- \frac \pi 2 - \alpha_0 \right) +\tan\theta \right), &  \theta < 0;
     \\ 
    z_s \left( \frac 1 2 \frac \xi {\cos^2\theta} \left(\frac 3 2 \pi - \alpha_0 \right) +\tan\theta \right),  &  \theta \ge 0,
  \end{cases}
\]
with corresponding group travel times:
\[
t_{g,  surf}(\theta)=\begin{cases}
     \frac{\omega m}{k_h g} \left( -\frac 1 2 \pi - \alpha_0 \right),  \theta < 0;
     \\ 
    \frac{\omega m}{k_h g} \left( \frac 3 2 \pi - \alpha_0 \right),  \theta \ge 0.
  \end{cases}
\]

Due to the symmetry of the initial conditions, the surface ripples will propagate circularly away from the source, following the time-distance relationship defined above using the take-off angle as parameter. Comparing these time-distance relations with the one for skip-distance, $\Delta= t_\Delta^2  g /{4\pi} m $ (see Appendix \ref{sec:appsolve} for details), it is useful to separately consider rays  going up and down from the source. The rays going up reach their upper turning point first and reflect back to the surface. After that the first surface ripple is defined by the rays leaving the source away from the surface and then reflected back. 
\begin{itemize}
\item when $z_s \geq z_E,$ after the up-propagating rays are reflected from the surface, the time-distance curve is similar to the skip-distance relationship with times slightly scaled, since the rays do not travel the whole skip-distance (distance between surface bounces). This scaling/difference depends on the source frequency and depth, via parameter $\xi$. See left column of Figure \ref{fig:wavefront} for an example of such source. 
\item when $z_s < z_E, $ after the initial surface bounce of "up" $(\theta < 0)$ rays, the time-distance curve is even closer to the skip-distance relationship, with wavefront from down rays slightly pre-ceding the skip-distance arrival time, while the up-ray wavefront which was initially reflected from the surface near the source, arriving slightly later due to the time it takes for the ray to travel from the source to the surface.
See right column of Figure \ref{fig:wavefront} for an example of such source. 
\end{itemize} 

%
%
%
%
%
%
%
%
%
%

\section{Discussion and applications to the Sun}\label{sec:discuss}

Before moving to the discussion of the applications of this method to the Sun, let us briefly review existing observations of the acoustic wave-fronts. While acoustic oscillations in the Sun have been extensively studied via global and local helioseismology, our understanding of the nature of their excitation remains relatively sketchy with stochastic turbulent convection generally considered to be the main mechanism responsible. There is, however, a recently discovered sunquake phenomena that provides us with observational examples of locally generated acoustic fields. Sunquakes, which are relatively rare events \citep[please see][for discussion on detectability]{donea11}, take place in active regions during flares, when the flare induced changes in sunspot penumbra under the right circumstances generate acoustic waves. These waves can be seen in photospheric Dopplergram observations as circular shaped surface ripples \citep{K2006,Moradi2007,Kosovichev2011} accelerating away from the source region. However, due to background oscillations, such ripples are often difficult to distinguish, so helioseismic methods such as time-distance diagram analysis and acoustic holography are usually applied to suitably processed series of line of sight velocity observations for quake detection. 

Time-distance method \citep{kz1998,Kosovichev2007,Kosovichev2011,ZGMZ2012} provides direct observational evidence of generated acoustic waves by integrating the signal over circles or arcs centred at the source for every observation in the time-series, thus producing a 2-D diagram displaying the field dependence on time and distance from the source. In such diagrams quake events have a clear signature in the form of a discernible time-distance ridge. It is the close agreement between the detected ridges and the theoretical travel-time skip-distance relation that provided the conclusive proof of sunquakes acoustic nature. On the other hand, the acoustic egression measurements \citep{Donea1999,DL2005,LD2008,MZZ2011} are based on theoretical modelling of acoustic waves propagating from a point source and provide a proxy measurement of acoustic energy emitted at a location at given time, thus producing a map of acoustic sources and sinks. Quake signatures in egression power maps are normally represented by a compact kernel or kernels of enhanced emission surrounded by the acoustically absorbing sunspot interior (see \citet{ZH2011,ZGMZ2012} for discussion and comparison of the two methods).

Both methods were originally based on the assumption of photospheric manifestation of a flare generated acoustic field in the form of circular shaped ripples propagating away from the source. However, considerable anisotropy in the acoustic amplitude of the ripples from the vantage of the sources has been observed for most quakes \citep{K2006,Moradi2007,donea11} with acoustic emission stronger in some directions than others and ripples changing shape becoming elliptical.\footnote{for example, using directional time-distance diagrams 
\cite{Kosovichev2011} has reported strong travel-time anisotropy in different arcs of the wave-front generated by the strongest seismic source of 2011 February 15 flare} In egression power maps such anisotropy is represented by ''stacked acoustic kernels'', where two or three narrow kernels are stacked together. 
\cite{Donea1999,DL2005} suggested that this could be the result of interference caused by the rapid motion of the source, roughly in the direction along which the kernels are stacked. This was supported by observations of apparent rapid photospheric movements at the quake locations. \citet{DL2005} found that for 2003 October 28 and 29 quakes the motion of the HXR sources was indeed aligned accordingly with egression power stacks. This was further confirmed for 2002 July 23 flare by \citet{Kosovichev2007} using HXR and Doppler data, where via data analysis and modelling the author estimated the speed of the seismic source to be supersonic\footnote{in the quiet Sun adiabatic soundspeed at the photospheric level is estimated around $7-8 {\rm \ km \ s}^{-1}$ in the quiet Sun} around $20-25 {\rm \ km \ s}^{-1}.$ More recently,  \citet{Kosovichev2011} used time-distance diagram analysis to detect a  supersonic movement of one of the seismic sources produced by 2011 February 15 flare, with speed around $15-17 {\rm \ km \ s}^{-1}$. Supersonic motions of around $14-22 {\rm \ km \ s}^{-1}$ have also been detected for this flare's second seismic source in \citet{ZGMZ2012}.
In the recent review \citet{donea11} notes that the maximum amplitude of the ripples emanating from a moving source is generally along the axis of the source, displaced from the source location in the direction of the motion. 

Both helioseismic methods have shown very local spatial nature of the seismic sources validating the point source assumption. However, there is no current consensus in regard to the physical mechanism and processes behind the quake excitation.  Back-warming heating, hydrodynamic shocks, particle precipitation and Lorentz force are the main scenarios currently considered and debated as those capable of producing flare acoustic response (see \citet[][]{donea11} and references therein for more details). Due to the lack of our current understanding of excitation, little is known about the depth of the source, but it is generally assumed to be near the photosphere since various modelling methods show only a small fraction of the energy initially invested in the shock penetrate into the photosphere.

\subsection{Stationary monochromatic source}
Back to the model considered in Sections \ref{sec:inicond}-\ref{sec:polysolve}, it is clear that non-evanescent acoustic waves are present in the model if and only if $k^2 \ge 0.$ Then the first condition on monochromatic source being able to generate acoustic wavefield is $\omega^2 > \omega_{ac}^2(z_s)$.  Let  $z_{cut}(\omega)$ denote the point such that $\omega_{ac}^2(z_{cut})= \omega^2$. Furthermore, as shown in Section \ref{app:ray_solve}, the presence of a unique extrema in the $k^2(z)=(\omega^2-\omega_{ac}^2)/{c^2}$ warrants the existence of the (upper and lower) turning point partition depth $z_E$. 
When such a partition depth exists, placing a source above it (but with $\omega > \omega_{ac}(z_s)$) will create a gap in ripples seen at the surface. Indeed, using the fact that for a fixed frequency the skip-distance for a particular ray is a function of its horizontal wavespeed, one can evaluate 
$$k_h^2 \leq \left. \frac{\omega^2-\omega_{ac}^2}{c^2} \right\arrowvert_{z=z_s}=k^2(z_s),$$ therefore $v_{ph}^2 \geq c^2 \omega^2/(\omega^2-\omega_{ac}^2). $ As $\omega_{ac}^2$ grows near the surface, $z$ approaches $z_{cut}$ and the phase speed grows infinitely large. 
As upper turning depths of the rays are greater than $z_E$, and since near upper turning points ray propagate nearly vertically, it is clear that rays going up from point source will not travel far from the source horizontally before being turned back down into the interior. For rays going down from the source the horizontal distance from the source to its first appearance at the surface will be slightly shorter than its skip-distance, so that the first appearance at the surface will be approximately at $\Delta(\min v_{ph}^2)$.
Thus, for example, for polytrope model considered in Section \ref{app:ray_solve}, taking $m=\frac 3 2$, at $z=122$ km only waves with frequency $\nu= \omega /{2\pi} \ge 7$ mHz can be generated. Moreover, given $z_s$, if $\omega$ is  close to the value of the $\omega_{ac}(z_s)$, the distance where generated waves surface away from the source can grow very large due to the fact that $k_h^2 \leq (\omega^2-\omega_{ac}^2) / {c^2}$. 

\begin{figure}
\includegraphics[width=8.5cm]{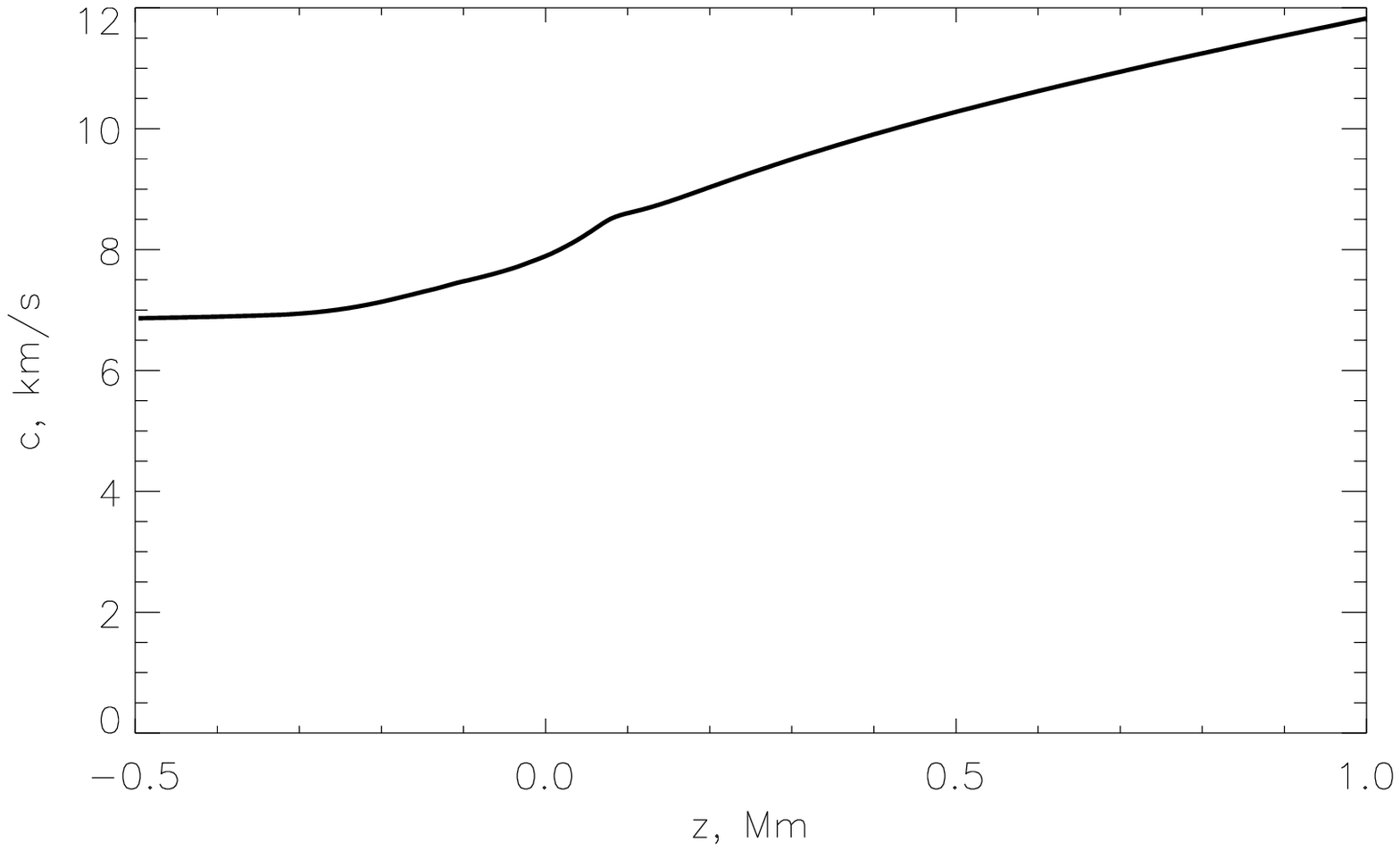} \\
\includegraphics[width=8.5cm]{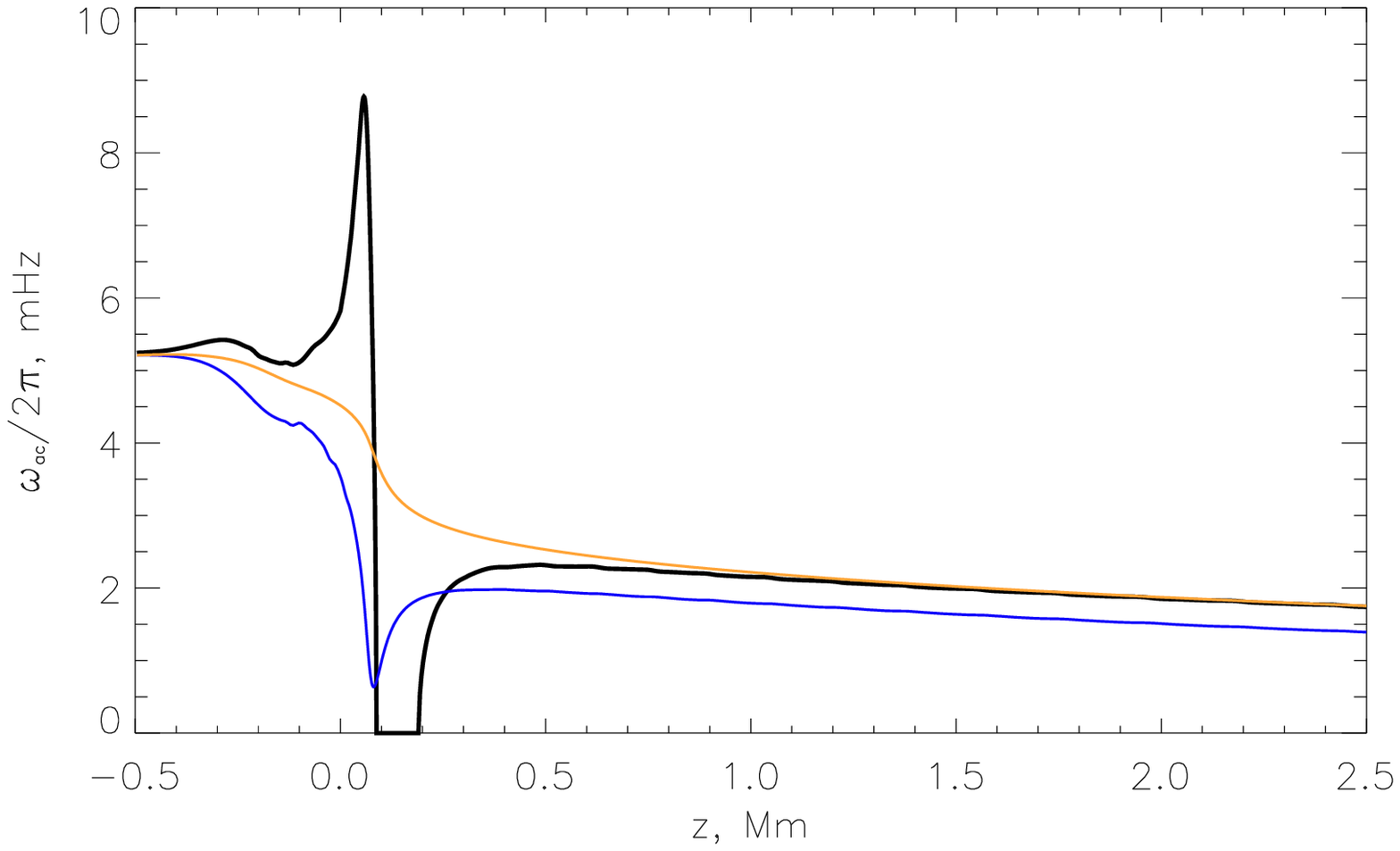}
\caption{Adiabatic soundspeed  ({\em left}) and acoustic cut-off frequency  ({\em right, black line}) from Christensen-Dalsgaard model C in the upper layers of solar interior. Note that the negative part of $\omega_{ac}^2$ has been truncated at zero. Yellow line is the isotermal cut-off frequency deduced from the model,  $\frac{c^2}{4H_p^2}$, blue line represents $\frac{c^2}{4H_\rho^2}$. 
\label{fig:modelC}}
\end{figure}

The assumption of a unique extrema for $k^2(z)$ doesn't hold in the case of 
Model C \citep{CDModel}, a widely used and thoroughly tested in helioseismology model of non-magnetic solar interior. This is due to multiple peaks in acoustic cut-off frequency present near the surface (see bottom panel of Figure \ref{fig:modelC}). This effect, however, is thought to be numeric in nature \citep[see][for example]{SchunkerCally} and other models such as isothermal cut-off frequency, ${c^2}/{4 H_p^2},$ are often used in the acoustic oscillation studies  in order to ensure smooth variation. It can be checked that in the isothermal case such an extrema exists, and corresponds to a maximum that appears to be unique below the surface for frequencies of up to $\approx 7$ mHz.

Adiabatic sound speed and acoustic cut-off frequency estimated from this model are presented in Figure \ref{fig:modelC}. It is seen from Figure \ref{fig:modelC}, where sound speed and acoustic cut-off frequency estimated from the model are plotted, that in the layers where $\omega_{ac}$ is large $c$ is of the order of 7-8 kilometres per second. Note that in the Sun high frequency rays will not necessarily have an upper turning point, escaping into the outer layers of solar atmosphere.

\subsubsection{Applicability of the method}
\label{sec:applicability}
Before going further, let us note some considerable limitations of the model and solution using the shallow source considered 
 in the
  earlier sections as an example. Stationary monochromatic source solution suggests that when the source is placed close to the surface the generated wavefront can surface some distance away from the source (right column, Figure \ref{fig:wavefront}). 
If this were to be applicable to sunquakes, it would imply that the detection of such events may require a field of view larger than currently used. And even then the detectability of such waves via local helioseismic methods might be difficult due to reduced wave amplitude at large distances from the source. 

However, clearly the model used here is too basic to expect a realistic representation of the physical phenomena. Apart from the glaring lack of magnetic field which we know is present at sunquake locations, the Brunt Vassala frequency is also zero, $N^2=0$ in (\ref{equ:main}). Employing ray-mechanical formalism \cite{BC2001} have shown the importance of Brunt Vassala term for waves near the acoustic cut-off boundary. 

There are also limitations associated with the solution itself. Firstly, by applying the condition $k_z^2 \ge 0$ and thus rejecting complex valued wave-numbers , the phenomena such as jacket modes is omitted, whereas those are needed to form a complete set of wave-equation solutions  \citep{BogdanCally1995}. This can possibly be addressed via extending the solution to complex rays, and for example considering the signal from downward propagating evanescent modes, \ie with imaginary $k_z$, and evaluating if and how it can transform into pure acoustic rays at larger depths.

Furthermore, the dispersion relation, corresponding Hamiltonian and rays associated with initial conditions are derived from the eikonal equation (\ref{eq:eik0}). The actual wave-field is then reconstructed by solving the set of transport equations (\ref{eq:transport}, \ref{eq:transport2}, ...) and postulating or establishing that the resulting series (\ref{equ:wkbansatz}) are asymptotic to the exact solution. Under certain conditions such as slow-variability of the model  \citep{KravtsovOrlov,gough2007}, 
an ''almost'' plane-wave (or eikonal)
 approximation can be justified, where the resulting field is approximated via its zero-order amplitude. The usual interpretation of rays as paths along which the energy propagates is based on such an approximation. This is clearly not the case for the shallow source where, due to the properties of acoustic cut-off frequency, model variability becomes high. 
As sufficient conditions for the 
''almost'' plane wave
approximation are clearly not met, this needs further investigation, e.g. solving higher order transport equations or approaching the problem via different methods.. In fact, based on more realistic wave modelling, there are reasons to think that physical situation is different.  (C.Lindsey - private communication).

Thus, the properties of the eikonal solution discussed here and in the following section should be considered as hypothetical, possibly posing questions about real physical scenarios and applicability of the geometric asymptotics method.

\input{mv_source}

%
%
%
%
%
%
%
%
%
%

\section{Conclusions}\label{conc} 

While the linear wave-equation has been extensively studied using geometric optics \citep[see][and references therein, for example]{KravtsovOrlov,KravstovOrlovCaustics}, the non-linearity induced by the presence of acoustic-cutoff frequency in the Sun has often been overlooked. Using a point source as an example, in this paper I have shown how suitable description of initial conditions can be used to derive geometric properties of acoustic fields governed by such non-linear wave-equation. 
For stationary monochromatic source a full solution of the eikonal equation including Jacobian computation and phase function reconstruction was derived, showing dependency of the field shape on source depth and frequency. The general treatment of a moving source has also been outlined, analytically deriving some properties that might be associated with sun-quakes. 
While primarily concentrating on mathematical treatment to the problem, some intriguing properties of phase functions corresponding to near surface sources have been deduced. It was shown that due to the presence of acoustic cutoff frequency a shock travelling with supersonic speed can generate a cone-like wave-packet in the direction of the shock movement.

Given the asymptotic nature of the method used, the paper poses interesting and current questions about acoustic waves generated near the solar surface. The resolution of such questions will also shed further light on limits of applicability of the geometric asymptotics in solar applications.

The availability of the full solution including the Jacobian suggests a possibility of using it to compute the Green's function via solving of zero-order transport equation. This, however, is complicated by the presence of the caustics, which require separate treatment and classification as outlined in \citet{KravtsovOrlov,KravstovOrlovCaustics}. While this is a subject of future study, other methods such as outlined in \citet[and references therein]{PH2010,LBH2011} might be better employed for this purpose.

Nonetheless, the method presented in this paper is shown to be a powerful tool for studying the behaviour of the wavefields when full initial conditions can be specified.

\section*{Acknowledgments}
I thank the referee, Dr Charles Lindsey, for his critical reading of the original version of the paper and many useful comments and suggestions that improved the presentation of the paper. 
I also acknowledge the Leverhulme Trust for funding the ÒProbing the Sun: inside and outÓ project upon which this research is based.
\bibliographystyle{mn2e1}
\bibliography{mnras-4rays}

\input{appendix}

\bsp

\label{lastpage}

\end{document}

%% file: method.tex
\section{Equation for scaled Pressure Perturbation and Geometrical Asymptotics}
\label{s:method}
Solar wave equation for scaled Lagrangian pressure perturbation, $\Psi=\rho_0^{-\frac 1 2} \delta p$, can be written \citep{gough93} as
\be
\frac{1}{c^2}\ddp{2}{}{t}\left( \ddp{2}{}{t} + \omega_{ac}^2 \right) {\Psi}-
\ddp{2}{}{t} \bigtriangledown^2 \Psi -N^2\bigtriangledown_h^2 \Psi =0. \label{equ:main}
\ee
Here $\btd_h^2$ is the horizontal Laplacian operator, $\rho_0$ unperturbed quiet sun density, $c^2$ adiabatic sound speed, $N^2$ Brunt-Vaasala buoyancy frequency,  and 
\be 
\omega_{ac}^2=\frac{c^2}{4H_\rho^2}\left(1- 2 \vect n . \btd H_\rho\right) \label{equ:om_ac}
\ee 
is Lamb's acoustic cut-off frequency, defined in terms of density scale height $H_\rho$ and unit vector $\vect n$ in the direction of the gravity action.  As is standard in acoustic mode helioseismology let us take $N^2=0$ to simplify the equation to a non-linear second-order wave-equation:
\be
\label{eq:waveLagrPress}
\frac{1}{c^2}\left( \ddp{2}{}{t} + \omega_{ac}^2 \right) {\Psi}-
 \bigtriangledown^2 \Psi=0
\ee

To find asymptotic solutions the method of geometrical optics \citep{KravtsovOrlov} is used in this paper. The method, developed for linear wave-equation and generalised as method of Geometric Asymptotics by \cite{GeoAsymptotics} for a wider range of high order differential operators. In short, this works by looking for the solution of the equation  (\ref{eq:waveLagrPress}) in terms of Debye ray series (or WKB ansatz \citet{Chapman99}) written for asymptotically large parameter $\Lambda$:
\be
\Psi(\vect r, t)=\sum_m^\infty \frac{A_m(\vect r, t)}{(i\Lambda)^m} e^{i\Lambda \varphi(\vect r, t)}.
\label{equ:wkbansatz}
\ee
On substitution of the series into  (\ref{eq:waveLagrPress}) and arranging the result in terms powers of the large asymptotic parameter $\Lambda$ a series of equations is obtained:
\begin{subequations}
\begin{align}
\left| \btd \varphi \right|^2-\frac 1{c^2}\left|\dd\varphi t \right|^2+\frac{\omega_{ac}^2}{c^2}& =0
\label{eq:eik0} \\
2\btd A_0 .\btd \varphi + A_0 \btd^2\varphi & - \nonumber
\\
- \frac 2 {c^2} \dd \varphi t \dd {A_0}t 
- \frac 1 {c^2} A_0 \ddp 2 \varphi t & = 0
\label{eq:transport} \\
2\btd A_1 .\btd \varphi + A_1 \btd^2\varphi & - \nonumber
\\
- \frac 2 {c^2} \dd \varphi t \dd {A_1}t 
- \frac 1 {c^2} A_1 \ddp 2 \varphi t & = -\left(\btd^2 -\frac 1 {c^2} \ddp 2 {} t \right) A_0 \label{eq:transport2} \\
\ldots \nonumber
\end{align}
\end{subequations}

Here (\ref{eq:eik0}) is a first order differential equation in partial derivatives called an eikonal equation which under suitable initial conditions can be solved to fully reconstruct phase function $\varphi(\vect r, t)$, 
 which is  also known as eikonal.
 The following (\ref{eq:transport}-\ref{eq:transport2}) are called transport equations and can be solved iteratively using the solution for eikonal to obtain amplitude coefficients $A_0, A_1, \ldots$. For more details and examples of the application of the method in the solar case see \cite{gough93}.

Let $M$ be a space-time manifold, described by Cartesian coordinates $(x, y, z)$ and time, $t$. Let $z$ be the depth, and $x$ and $y$ correspond to horizontal directions. Then, as $\varphi=\varphi(x, y, z, t)$, the following variables describe $\varphi$ in the phase space, $\Omega M$
\be 
k_x=\dd \varphi x , \  k_y=\dd \varphi y,  \  k_z=\dd \varphi z, \ \omega= - \dd \varphi t. \label{equ:ini_phase}
\ee
The eikonal (\ref{eq:eik0}) belongs to the class of Hamilton-Jacobi equations and can be solved by method of characteristics in the phase space. In the above notation it corresponds to the Hamiltonian 
\be
H=\frac 1 2 \left( k_x^2+k_y^2+k_z^2-\frac{\omega^2-\omega_{ac}^2}{c^2} \right),
\label{eq:Hamiltonian3D}
\ee
so that the equation (\ref{eq:eik0}) becomes $H=0$ and finding the solution is reduced to integration of the characteristic system.

It can be shown (see Section \ref{s:ini_mono_spherical} for example) that when the Hamiltonian (\ref{eq:Hamiltonian3D}) does not explicitly depend on horizontal variables, under suitably symmetric initial conditions, $M$ can be reduced by one dimension, so that the solution can primarily be sought in two spatial dimensions, i.e. $\varphi=\varphi(x, z, t)$, so that
\begin{subequations}
\begin{align}
H=\frac 1 2 \left( k_h^2+k_z^2-\frac{\omega^2-\omega_{ac}^2}{c^2} \right),  
\label{eq:Hamiltonian} \\
k_h=\dd \varphi x,  \  k_z=\dd \varphi z, \ \omega= - \dd \varphi t. 
\label{eq:Hamiltonian_defs}
\end{align}
\end{subequations}
The full solution is then obtained via rotation using $k_h^2=k_x^2+k_y^2$ relationship. 

In the problems considered in this paper, the sound speed and acoustic cut-off frequency depend only on depth, so the time-dependent equation (\ref{eq:waveLagrPress}) is often reduced to
two spatial dimensions plus time. In such case we introduce cartesian coordinates $x$ and $z$ on manifold $M,$ corresponding to horizontal and vertical directions, with depth $z$ chosen to be 0 at the surface and positive below it (see Figure \ref{fig:geo_setup}). 
Note that in six dimensional space $\Omega M$ the graph of $\varphi$ (\ie the map $(x, z, t) \mapsto \left(x, z, t, \dd \varphi x, \dd \varphi z, \dd \varphi t \right)$) is a three-dimensional surface. The characteristic system is then defined as
\begin{subequations}\label{eq:HJsystem}
\begin{empheq}[left=\empheqlbrace]{align}
\label{eq:HJsystem0}
\dder x \tau & = \dd H{k_h}, \ \ 
\dder z \tau  = \dd H{k_z}, \ \  \dder t \tau = -\dd H \omega, \\
\dder {k_h} \tau & = -\dd H x, \ \ 
\dder {k_z} \tau = -\dd H z, \ \ \ \dder{\omega}\tau= \dd H t, \label{eq:HJsystem1}
\end{empheq}
\end{subequations}
where $\tau$ is an independent variable. Note that  condition $H=0$  can be used instead of one of the equations in the above system. Solutions of the system 
(\ref{eq:HJsystem}) 
written in terms of $\tau$ (\ie $x(\tau), z(\tau), t(\tau), k_h(\tau), k_z(\tau), \omega(\tau)$) provide us with characteristic lines lying on the graph of $\varphi$ in $\Omega M$. The phase function is determined by integrating
\be
\dder \varphi \tau = k_h \dder x \tau + k_z \dder z \tau - \omega \dder t \tau \label{eq:phase_d}
\ee
 along each such line. In order to reconstruct the phase function $\varphi$ in the whole  $(x, z, t)$ domain, initial conditions on some two-dimensional surface $S \in \Omega M$ have to be provided so that characteristic lines originating from every point of such surface sweep out the three-dimensional graph. Let $(\zeta, \eta)$ be the coordinates on $S$ so that the initial field can be written as $$\Psi_0(\zeta, \eta)= e^{i\Lambda \varphi(\zeta, \eta)} \sum \frac{A_m^0(\zeta, \eta)}{(i\Lambda)^m}.$$ Characteristic solutions of (\ref{eq:HJsystem}) with initial conditions taken as described sweep out a three-dimensional surface in $\Omega M$ described by $(\zeta, \eta, \tau)$ which are called ray-coordinates. Thus, it is important to take into account the proper initial conditions in order to understand the geometrical aspects of the wave-field associated with the equation  (\ref{eq:eik0}).  Projections of individual characteristics  $\left(x(\tau), z(\tau), t(\tau), k_h(\tau), k_z(\tau), \omega(\tau) \right)$ from $\Omega M$ onto $(x, z, t)$ space are called rays. 

Once the eikonal equation solved, its solution can then be used to reconstruct amplitude along the rays by rewriting the transport equations in ray coordinates. These take form:
\begin{subequations}
\begin{align}
 \dder{A_0^2}{\tau}+A_0^2 
 \dder {\ln \mathcal D(\tau)}\tau
& = 0,
\label{equ:transport0} \\
 2 \dder {A_1}\tau + A_1 \dder {\ln \mathcal D(\tau)}\tau  & = -\left(\btd^2 A_0 - \frac 1 {c^2}\ddp 2 {A_0} t \right), 
\label{equ:transport1} \\
 \ldots & \nonumber
 \end{align}
\end{subequations}

Here $\mathcal D(\tau)=\dd {(x, z, t)}{(\zeta, \eta, \tau)}$ is the Jacobian of the transform from cartesian to ray coordinates. Zero-order equation has the solution \citep{KravtsovOrlov}: 
\be
A_0(\zeta, \eta, \tau)=A_0^0(\zeta, \eta) \left[ \frac{\mathcal D (\tau_0)}{\mathcal D(\tau)}\right]^{\frac 1 2}.
\label{eq:trans0_sol}
\ee
Note that $A_0$ depends only on geometry of the problem and the initial values of the field along the ray, thus expressing the local character of zero-order WKB approximation. So for example, in order to reconstruct the field at some point $(x_1, z_1, t_1)$, one normally determines rays coordinates corresponding to that point via ekional equation solution and then uses (\ref{eq:trans0_sol}) to find the value of $A_0$ and (\ref{eq:phase_d}) to find the phase-function. 

Higher order coefficients can be written as 
\begin{empheq}{align}
A_m  = & A_m^0  \left[ \frac{\mathcal D (\tau_0)}{\mathcal D(\tau)}\right]^{\frac 1 2}   -\frac 1{2\sqrt{\mathcal D(\tau)}} \times \nonumber
\\
 & \times \int_{\tau_0}^\tau \left(\btd^2 A_{m-1} - \frac 1 {c^2}\ddp 2 {A_{m-1}} t \right) \sqrt{\mathcal D(\tau)} d\tau'.
\end{empheq}
The  presence of the Laplacian and higher order time derivative in the integral on the right hand side in the above means that $A_m$ depends not only on the values of $A_{m-1}$ immediately on the ray, but also the values $A_{m-1}$ in ray vicinity. Thus higher order amplitudes describe diffraction effects discarded in zero-order approximation.

At points where the Jacobian is zero, $A_0$ and higher order coefficients grow infinitely large, thus defining regions where the method breaks down. Such regions require separate analysis such as matching asymptotics or other methods \citep{KravtsovOrlov,KravstovOrlovCaustics}. Points where the Jacobian is zero are said to form caustic surfaces, or caustics. Such surfaces envelope ray surfaces and are singularities of projection of the graph of $\varphi(\zeta, \eta, \tau)$ from $\Omega M$ to $(x, z, t)$ space \citep{ArnoldBook}.

It is  common in many applications to disregard the non-zero order terms in (\ref{equ:wkbansatz}), $A_m=0, \forall m>0,$ thus 
assuming the wave-equation (\ref{eq:waveLagrPress}) solution to be an ''eikonal approximation''  $\Psi\approx A_0 \exp (i\Lambda \varphi),$ also known as an ''almost plane wave".
\citet{KravtsovOrlov,gough2007} give necessary conditions for this as that the medium, $c^2$ and $\omega^2_{ac}$ varies slowly over the characteristic length-scale of variation of the wave.
 

%% file: initial_conditions.tex
\section{Initial conditions} \label{sec:inicond}
\subsection{General approach}
In $3+1$ dimensions consider initial surface $S \subset M$ parametrised by some coordinate set $(\theta, \phi, t_0),$ with $M$ described by $(x, y, z, t)$. Then the field on such surface can  be written as 
\be
\Psi_0(\theta, \phi, t_0)=A_0(\theta, \phi, t_0) e^{i \varphi_0 (\theta, \phi, t_0)}. \label{eq:inifield}
\ee
Solution of the 3D characteristic system 
\begin{subequations} \label{equ:genHamiltonian}
\begin{empheq}[left=\empheqlbrace]{align}
\dder{\vect x}{\tau} & =\dd H {\vect{k}}, \label{equ:genHamiltonian0} \\
\dder{\vect k}{\tau} & = - \dd H {\vect{x}}, \label{equ:genHamiltonian1}
\end{empheq}
\end{subequations}
deduced from the Hamiltonian (\ref{eq:Hamiltonian3D}), provides  the graph of $\varphi,$ 
$$\left(x, y, z, t, \dd \varphi x,  \dd \varphi y,  \dd \varphi z,  \dd \varphi t  \right ) \subset \Omega M,$$ 
as function of ray-coordinates $(\theta , \phi, t_0, \tau).$ Here $\tau$ is the parameter along each ray, $\tau = 0$ on $S$, and $\theta, \phi$ and $t_0$ are ray-numbering coordinates.  As $S$ is submerged in $M,$ there is a natural mapping, $(\theta, \phi, t_0) \hookrightarrow (x, y, z, t),$ providing the initial conditions for (\ref{equ:genHamiltonian0}). To recover the initial conditions for momentum part (\ref{equ:genHamiltonian0}) of the characteristic system, recall the definition (\ref{equ:ini_phase}) of the momentum variables, and let $k_{x0}, k_{y0}, k_{z0}, \omega_0$ be their values on the initial surface $S$. It then follows that
\begin{subequations} \label{equ:ini3d}
\begin{empheq}[left=\empheqlbrace]{align}
\dd {\varphi_0} \theta & = k_{x0} \dd x \theta +  k_{y0}\dd y \theta + k_{z0} \dd z \theta - \omega_0 \dd t \theta, \label{equ:ini3d1} \\
\dd {\varphi_0} \phi & = k_{x0} \dd x \phi +  k_{y0}\dd y \phi + k_{z0} \dd z \phi - \omega_0 \dd t \phi,   \label{equ:ini3d2}\\
\dd {\varphi_0} {t_0} & = k_{x0} \dd x {t_0} +  k_{y0}\dd y {t_0} + k_{z0} \dd z {t_0} - \omega_0 \dd t {t_0}, \label{equ:ini3d3}
\end{empheq}
\end{subequations}
which together with 3-D Hamiltonian (\ref{eq:Hamiltonian3D}) evaluated at point $(\theta, \phi, t_0, k_{x0},  k_{y0}, k_{z0}, \omega_0)$ forms a set of four equations for the four unknowns ($k_{x0},  k_{y0}, k_{z0}, \omega_0$). These are then solved in terms of ray-numbering coordinates $(\theta, \phi, t_0),$ thus recovering full initial conditions for the characteristic system.  Solution of the system (\ref{equ:genHamiltonian}) provides the map $$ \left( \theta, \phi, t_0, \tau \right)  \mapsto \left( x, y, z, t, k_x, k_y, k_z, \omega \right) \in \Omega M,$$ sweeping out the graph of the phase function $\varphi(x, y, z, t)$ in the phase space from the initial surface $S$. The phase function itself and amplitude coefficients are then reconstructed as described above.

%
%
%
%
%
%
%
%
%
%

\subsection{Monochromatic spherical point source}
\label{s:ini_mono_spherical}
Let us now derive the initial conditions for a spherical source. Consider a static sphere of a small radius $r$ located at depth $z=z_s$ to be the surface $S$ where the initial conditions for pressure perturbation are set. The surface can be parametrised by the spherical coordinates plus time, $(\theta, \phi, t_0)$, so that the Cartesian coordinates and time are expressed as
\begin{subequations} \label{eq:ini_sc3D}
\begin{empheq}[left=\empheqlbrace]{align}
x &=x_s+r \cos \theta \sin \phi \label{eq:ini_sc3D1} \\
y &=y_s+r \cos \theta \cos \phi \label{eq:ini_sc3D2} \\
z & =z_s+r \sin \theta \label{eq:ini_sc3D3} \\
t & =t_0, \label{eq:ini_sc3D4}
\end{empheq}
\end{subequations}
where $\phi \in [0, 2 \pi), \theta \in \left[-\frac \pi 2, \frac \pi 2\right].$
Using (\ref{eq:inifield}-\ref{equ:ini3d}), let $k_{h0}^2=k_{x0}^2+k_{y0}^2$ and define $k_{\phi 0}$ so that $k_{x0}=k_{h0}\cos k_{\phi 0}, k_{y0}=k_{h0} \sin k_{\phi 0}$,  system (\ref{equ:ini3d}) then simplifies to 
\begin{subequations} \label{eq:ini_dvals}
\begin{empheq}[left=\empheqlbrace]{align}
\dd {\varphi_0} \theta & = k_{z0} r \cos \theta - k_{h0} r \sin \theta \sin \left(k_{\phi 0}+\phi\right), \label{equ:ini_dtheta} \\
\dd {\varphi_0} \phi & =  k_{h0} r \cos \theta \cos \left(k_{\phi 0}+\phi\right), \ \label{equ:ini_dphi} \\
\dd {\varphi_0} {t_0} & = - \omega_0.  \label{equ:ini_dt0}
\end{empheq}
\end{subequations} 
When sound speed and acoustic cut-off frequency depend only on depth, Hamiltonian (\ref{eq:Hamiltonian3D}) is explicitly independent of $x, y$ and $t$, so from (\ref{equ:genHamiltonian0}-\ref{equ:genHamiltonian1}) it follows that $k_x, k_y$ and $\omega$ are constant along each ray. Therefore, $k_h$ and $k_\phi$ extended from the above definition throughout the phase space are also  constant along rays. When, in addition, the initial phase function $\varphi_0$ does not depend on $\phi,$ e.g. $\ddil {\varphi_0} \phi = 0$, it follows that $\cos \left(k_{\phi 0}+\phi\right)=0$, so that via a suitable choice of the $x$- and $y$-axis directions $k_{\phi 0}+\phi = \frac \pi 2 + 2 \pi n.$ In this case the system can be reduced to two-dimensions plus time  by letting $\phi=\frac \pi 2, k_{\phi 0}=0$ in (\ref{eq:ini_sc3D}), introducing $\hat x = x- x_s,$ and solving system (\ref{eq:HJsystem}),
i.e. using $(\theta, t_0)$ to find $(\hat{x}, z, t, k_h, k_z, \omega).$  Using symmetry, this solution 
is then rotated by varying $\phi, k_\phi$ to obtain the full three-dimensional field:
$$
(\theta, t_0, \tau) \mapsto (\hat{x}, z, t, k_h, k_z, \omega)  \hookrightarrow (x, y, z, t, k_x, k_y, k_z, \omega),
$$
via
\begin{subequations}\label{eq:rotation_transform}
\begin{align}
x & =  x_s+\hat x \sin \phi,  y  = y_s+\hat x \cos \phi, \label{eq:rotation_transform1} \\
k_x&  =  k_h \sin \phi,  k_y  =  k_h \cos \phi. \label{eq:rotation_transform2}
\end{align}
\end{subequations}
Point source is approximated by letting $r$ in the definition (\ref{eq:ini_sc3D}) to be infinitely small. A source is spatially homogenous when the initial phase function, $\varphi_0$ in (\ref{eq:inifield}), may depend only on time.

When the dimensionality is reduced, surface $S,$ where initial conditions for the pressure perturbation are set, becomes a circle of radius, $r$, located at depth $z=z_s.$ The surface is parametrised by coordinates $(\theta, t_0)$ with 
\begin{subequations} \label{eq:ini_sc}
\begin{empheq}[left=\empheqlbrace]{align}
x & =x_s+r \cos \theta \label{eq:ini_sc1} \\
z & =z_s+r \sin \theta \label{eq:ini_sc2} \\
t & = t_0 \label{eq:ini_sc3}
\end{empheq}
\end{subequations}
The initial field on this surface is described as 
\be
\Psi_0=A_0(\theta, t_0) e^{i \varphi_0 (\theta, t_0)}. \label{eq:ini}
\ee
The unknown initial wavenumbers and frequency,  $k_{h0}=\ddil {\varphi_0}{x}, k_{z0}= \ddil {\varphi_0}{z}$ and $\omega_0=-\ddil{\varphi_0} t$, are found from the system:
\begin{subequations} \label{eq:dg_varphi}
\begin{empheq}[left=\empheqlbrace]{align}
\dd{\varphi_0}{t_0} & =k_{h0} \dd x {t_0} + k_{z0} \dd z {t_0} - \omega_0 \dd t {t_0},
\label{eq:dg_varphi1} \\
\dd{\varphi_0}{\theta} & =k_{h0} \dd x \theta + k_{z0} \dd z \theta - \omega_0 \dd t \theta, 
\label{eq:dg_varphi2} \\
k_{h0}^2+k_{z0}^2 & =\frac{\omega_0^2-\omega_{ac}^2}{c^2}.
\label{eq:dg_varphi3}
\end{empheq}
\end{subequations}
Let us now consider spherical monochromatic homogeneous source of some fixed frequency, $\omega_f $, \ie $\varphi_0(\theta, t_0)=-\omega_f t_0.$ Then from the above $\omega_0=\omega_f$ and $k_{h0}\sin \theta = k_{z0}\cos \theta.$ This implies that in this configuration (see Figure \ref{fig:geo_setup}) $\theta$ can be viewed as ray take-off angle and all rays generated from $S$ are of the same frequency, so frequency subscripts can be dropped. Define $$k_s^2(z_s, \omega)=\left.\frac{\omega^2-\omega_{ac}^2}{c^2} \right\arrowvert_{z=z_s},$$ then horizontal and vertical wavenumbers can be rewritten as $k_{h0}=k_s \cos \theta,$ and $k_{z0}=k_s \sin \theta.$  Since Hamiltonian is independent of horizontal coordinate and time, $\omega$ and $k_h$ are constant on each ray. Then so is the horizontal phasespeed of a ray, $$v_{ph}^2=\frac {\omega^2}{k_h^2}= {\frac{c^2}{\cos^2\theta}\frac 1{1-\frac{\omega_{ac}^2}{\omega^2}}}.$$

When $t_0$ is fixed, $\theta$ serves as a ray-numbering parameter. Note that by setting $A(\theta, t_0)=0$, when $\theta < 0$ one can investigate the wavefield generated by rays going down from the source, which would correspond to semi-spheric pump creating pressure at the source. Similarly, taking $A(\theta, t_0)=0$ for $\theta > 0$ corresponds to the wavefield generated by rays going up to the surface.


%% file: mv_source.tex
%
%
%
%
%
%
%
%
%
%

\subsection{Moving source}
\label{sec:mvnsrc}

Let us now consider a moving source. This is  achieved by going back to (\ref{eq:ini_sc3D}) and letting the source coordinates $\left(x_s, y_s, z_s \right)$ depend on time $t_0.$ It is clear that the symmetry is broken unless the source moves along the axis of rotation, so the three-dimensional equations are considered. When solving the system (\ref{equ:ini3d}), use the stationary source case outlined in Section \ref{s:ini_mono_spherical}. As $\ddil {\varphi_0}{\theta}$ and $\ddil {\varphi_0}{\phi}$ are actually the same, $k_{h0}$ and $k_{\phi 0}$ are defined in the similar manner, and with equations (\ref{equ:ini_dtheta}-\ref{equ:ini_dphi}), it also follows that  $\cos \left(k_{\phi 0}+\phi\right)=0.$ It is only the frequency definition (\ref{equ:ini_dt0}) that is affected by source movement: 
$$\dd {\varphi_0}{t_0}= {\vect k}_0 . \vect v - \omega_0,
{\rm \ \ where\ \ } \vect v=\dder {{\vect x}_s} {t_0},$$ 
represents source velocity. Similar to Section \ref{s:ini_mono_spherical} let the initial phase in the definition (\ref{eq:ini}) be monochromatic and homogenous, \ie $\varphi_0= -\omega_f t_0.$ Then, the initial ray frequency is given as 
\be
\omega_0=\omega_f +  {\vect k}_0 . \vect v . \label{equ:freq_modified}
\ee
Hence, in this case, the source frequency is modified depending on the wave-vector: ray frequency which is constant along each ray  varies from one ray to another according to $\omega_0=\omega_f+k_0 v \cos \gamma$, where $\gamma$ is the angle between the wave vector and source velocity. The frequency increases along the vector of the source movement, and decreases when the wave-vector points in the opposite direction. 

Similar to stationary source (see Section \ref{s:ini_mono_spherical}), it follows from the first two equations in \eqref{equ:ini3d} that $k_h=k_0\cos \theta, k_{z0}=k_0 \sin \theta,$ where $k_0^2=({\omega_0^2-\omega_{ac}^2})/{c^2}$. Then use the dispersion relation $H=0,$ where Hamiltonian $H$ is defined in (\ref{eq:Hamiltonian3D}), to find $k_0$.  Substituting (\ref{equ:freq_modified}) into the last expression, the following quadratic equation with respect to $k_0$ is obtained:
\be
k_0^2(u^2-1)+2\frac{\omega_f}{c}u k_0 + \frac{\omega^2_f-\omega_{ac}^2}{c^2}=0,
\ee
where $u=v \cos (\gamma  / c).$ Consider the case when, as before, the source with frequency greater than the acoustic cut-off value, $\omega_f \geq \omega_{ac}(z_s(t_0))$ moves with subsonic speed, $v<c \Longrightarrow u^2 < 1.$ 
The solution is 
$$
k_{0\pm}=\frac 1 c \frac {\omega_f u  \pm \sqrt D}{1-u^2} \Longrightarrow \omega_{0\pm}=\frac 1{1-u^2}\left(\omega_f \pm u \sqrt D\right),
$$ 
where $D = \omega_f^2 - \omega_{ac}^2+\omega_{ac}^2 u^2.$ Clearly, under our assumptions $D>0.$
Since $k_0$ is defined as length of the wave-vector, we are looking for positively valued roots only, which leaves only $k_{0+}$ and $\omega_{0+}.$

To visualise the nature of the angle $\gamma$ let us now define an appropriate coordinate frame. Assume that at the time $t_0$ the source is located at $x_s(t_0)=y_s(t_0)=0.$ Since there is freedom in selecting the direction of $x$- and $y$- axes, choose coordinate axes so that the source moves in the direction of $y$-axis, \ie $\vect v = (v_x, 0, v_z).$ In spherical coordinates, $(r, \theta, \phi),$ defined for the source, $\vect v=(v, \lambda, 0),$ where $\lambda$ is the angle between the velocity vector and $xy$-plane. Then, using the spherical law of cosines, for any wave-vector ${\vect k}_0=(k_0, \theta, \phi)$ we have
\be
\cos \gamma = \sin \theta \sin \lambda + \cos \theta \cos \lambda \cos \phi. \label{equ:angle3d}
\ee
Note that apart from the assumption of dependence on depth only all of the above does not depend on the model for the sound speed and acoustic cut-off.

In the polytrope model the ray solution derived in the Appendix, (\ref{eq:alpha_x}-\ref{eq:alpha_alpha0}) together with (\ref{eq:alpha_0_sol}) and (\ref{eq:sol_phase_var}),  holds for individual rays. However, the generated wave-field will have a more complex dependence on the parameter $t_0,$ due to the source movement. E.g. at moment $t_0$ source produces a family rays governed by the above equations,  then at moment $t_0+\delta t$ the source will have moved producing another family of rays. This will be investigated in future publication.

\subsection{Supersonic source}
\label{s:movingsource}
While the wave-fields generated by a monochromatic point source have some of the properties associated with solar quakes, one of the main limitations of the considerations above is the assumption of source frequency. Even for observed high frequency\footnote{sunquake egression signal is usually strong around $\nu=6 {\rm \ mHz}$}  waves generated by flares, the period is still of  the order of couple of minutes. At the same time, the photospheric changes observed at the quake locations happen on a much shorter scale  \citep{ZZ2007,Kosovichev2007,Kosovichev2011,ZGMZ2011,ZGMZ2012}.  

It is instructive to consider a non-harmonic source, \ie $\varphi_0(\theta, \phi, t_0)=0.$ This corresponds to the wave-field on initial surface simply written as   $\Psi_0=A_0(\theta, \phi, t_0),$ meaning that the pressure perturbation, $\rho^{\frac 1 2}\delta p,$ is varying slowly on the initial surface. 
As in previous Section \ref{sec:mvnsrc},  $\ddil {\varphi_0}\theta = 0$, $k_{z0}=k_{h0} \tan \theta$, system \eqref{equ:ini3d} leads to the following solution:
\begin{subequations}
\begin{empheq}[left=\empheqlbrace]{align}
k^2_0 & =\frac{\omega_{ac}^2}{v^2 \cos^2 \gamma -c^2}, \label{eq:ms_k0}\\
\omega_0 & =\vect{k}_0.\vect{v} = k_0 v \cos \gamma, \label{eq:ms_freq}\\
k_{h0} & =k_0 \cos \theta, \ \  k_{z0}=k_0 \sin \theta, \label{eq:ms_waven}
\end{empheq}
\end{subequations}
where $\vect k_0$ is the wavevector, $\vect v$ is source velocity, $v=|\vect v|,$ and $\gamma$ is the angle between the two vectors.
From the above it follows that non-evanescent acoustic waves ($k_0^2 > 0$) are generated if and only if the source moves with supersonic speed $v^2>c^2$. Moreover, let us rewrite (\ref{eq:ms_freq}) as 
\be
\omega_0=\omega_{ac} \frac{\cos \gamma}{\sqrt{\cos^2 \gamma-\frac{c^2}{v^2}}},
\label{eq:ms_omega_0}
\ee
where the square root is always taken with plus sign due to (\ref{eq:ms_k0}) and the fact that $k_0>0$ by definition. 
Again, it is clear that while the frequency is constant on each individual ray, it will vary from ray to ray as parameterised by $\theta, \phi$ and $t_0$. It is also evident that in this case $\omega_0$ is always greater then the value of $\omega_{ac}$ at the source depth, going to infinity as $\cos^2 \gamma \longrightarrow {c^2}/{v^2}<1$. In addition, $\omega_0 \geq \omega_{min}={\omega_{ac} v}/ ({\sqrt{v^2-c^2}})$. Note that when $\cos \gamma$ is negative $\omega_0$ becomes negative, so waves waves will only be generated in the direction of the source movement. Moreover, given the \eqref{equ:angle3d},  the condition 
\be
\cos^2 \gamma > \frac{c^2}{v^2} \label{eq:ms_theta_cond}
\ee
ensures that rays are generated for only a relatively narrow range of values of $\theta$ and $\phi$, essentially forming a cone around the velocity vector $\vect v$.
 The horizontal phase speed at the source does not depend on source depth: $v_{ph}= {\omega_0}/{k_{h0}}= v {\cos \gamma}/{\cos \theta} .$

Again, note that  apart from the assumption of dependence on depth only all considerations in this section are independent of the model of the media where the waves propagate and rely only the dispersion relation as well as spatial geometry of the problem. Also, note that this solution is only possible when $\omega_{ac}>0.$ 

\subsection{Comparison with sunquakes}
As quake observations and modelling suggest the source is located near surface, let us assume that the source moves in the upper ranges of solar interior, so that for generated frequencies $z_s < z_E$, i.e. the source is located near upper turning points. Then, for each generated ray, the horizontal distance from the source approximately equals to the skip-distance,  $\Delta(k_h, \omega)$. On the other hand, the skip-distance is essentially a function of horizontal phase speed of the ray with lower phase-speed values corresponding to smaller distances. The relationship is exact for polytrope and other theoretical models \citep[see][]{CDLN} and has been observationally validated by the time-distance helioseismology and acoustic holography.

Let us consider a source propagating with velocity $v$ vertically downward, \ie $\lambda=\frac \pi 2$ (see \eqref{equ:angle3d}).  Then $\cos \gamma = \sin \theta$, and as only downward propagating rays are generated, the first appearance of the generated wavefront on the surface corresponds to the minimum value of $v_{ph}=v \tan \theta$. Then for $\theta \in \left[ 0, \frac \pi 2 \right]$ the minimum horizontal phase speed will be achieved at the lowest value of $\theta$. Using (\ref{eq:ms_theta_cond}) one obtains
\be
\min v_{ph}=\frac{vc}{\sqrt{v^2-c^2}}, \label{eq:min_vph}
\ee
where expression on the right hand side is evaluated at $z=z_s$.
However, frequency of the rays travelling near such phase speed will be approaching infinity due to (\ref{eq:ms_omega_0}). Hence if observations are made at certain Nyquist frequency, $\omega_N,$ such waves may not be observed, so further restrictions need be considered, namely, $\omega_{0} \leq \omega_N$ has to hold. From \eqref{eq:ms_omega_0} it follows that this condition is equivalent to
%
%
\be
\left.
\sin^2\theta\geq \frac{c^2}{v^2}\frac{\omega_N^2}{\omega_N^2-\omega^2_{ac}}
\right\arrowvert_{z=z_s}
. \nonumber
\ee
From this the minimum ''observable'' phase-speed can be evaluated:
\be
\left.
\min v_{ph}^{obs}=\frac{vc}{\sqrt{\left(1-\frac{\omega_{ac}^2}{\omega_N^2}\right)v^2-c^2}}
\right\arrowvert_{z=z_s}
\ee
Therefore, in the acoustic wave-field excited by a vertical supersonic shock perturbation only waves with phase speed exceeding these will be observable. 
Moreover, as surface ripples from such source are determined by the phase speed, the minimum distance away from the source can be estimated using the above inequality. 
For example, if $c=8 {\rm km/s}, v=10 {\rm km/s}$, then minimum $v_{ph} \approx 13.3 {\rm km/s}.$ For the Sun this corresponds to a skip-distance of around $5-7$ Mm. 
Let us take the value of cut-off frequency at source depth, $\omega_{ac}(z_s)/2\pi,$ to be $\approx 5.$ mHz.
Then adding the condition that the cyclic frequency along the ray is no greater than $8.4$ mHz, gives us a minimum phase speed estimate at  $\approx 106  {\rm km/s}$. Therefore, in this case, only  ripples at distances of order of hundred megameters from the source would be potentially observable.

More generally, let us rewrite \eqref{equ:angle3d}: 
$$\cos \gamma = \cos (\theta - \lambda) - \cos \theta \cos \lambda ( 1 - \cos \phi).$$
Then, as $\cos \theta$ and $\cos \lambda$ are non-negative, the inequality $\cos \gamma >  c / v$ implies that $\cos(\theta - \lambda)  >  c/  v > 0.$ Hence,  $$\theta \in \left( \lambda - \arccos \frac c v, \lambda + \arccos \frac c v \right).$$ It is also clear $\cos \phi  >  c / v - \sin\theta \sin \lambda.$  Thus for downward propagating source, \ie $\lambda \in \left[ 0, \frac \pi 2 \right],$ very roughly it can be estimated $\cos \phi  >  c / v - \sin \lambda,$ so only waves in the limited range of $\phi$ can be generated. But as rays propagate in the plane defined by $z$-axis and their initial wave-vector, this means that surface ripples will only appear in limited arcs in the direction of the source movement in this way producing anisotropy in wavefront amplitude. 

Given supersonic movements observed at the sunquake source locations, and strong anisotropy in amplitude in the direction of the movement, it is tempting to hypothesise that this mechanism plays an important role in the sun-quake wavefront generation. However, caution needs to be exercised as the problem has been approached here from purely mathematical standpoint and the notoriously difficult question of method applicability has not been addressed. 
While this solution appears to provide a possible explanation for some of the observed properties of the sun-quake generated acoustic fields, the applicability of the method under the circumstances should be questioned as the variations are obviously very fast and known ''sufficient'' conditions for method applicability do not hold in this case (see Section \ref{sec:applicability}). Numerical simulation or modal approach can provide a way to check this. It is interesting to note that published simulations showing generation of acoustic waves by convective vortices \citep[][]{Kitiashvili2011,Moll2011} also report supersonic movements at such vortices.
Clearly further analysis is required to address these issues as well as to investigate various dependencies that in this paper were touched upon only briefly.

{
The anisotropy of flare generated acoustic wave-fronts has so far been attributed to the effects of magnetic field and more general plasma subsurface properties in the Active Regions where quakes take place. It is indisputable that such conditions affect and likely shape generated waves. However, it can be argued that the effects discussed in this paper, if supported by more realistic analysis as well as for magnetised case, can play an important role in determining the shape and properties of the wave-field, which is then inadvertently affected by the magnetic field, flows and other inhomogeneities along its path. 

On the other hand, it can be seen that even without the magnetic field, \eg when moving source is considered, the number of variables on which solution depends grows quite fast. Given our understanding of interaction of acoustic and magneto-acoustic waves, \citep[see][and references therein]{SchunkerCally,Moradi2010}, it is clear that the inclusion of magnetic field effects is highly likely to paint an even more complex picture.}

%% file: appendix.tex
\begin{figure}
\centering
\includegraphics[width=8cm]{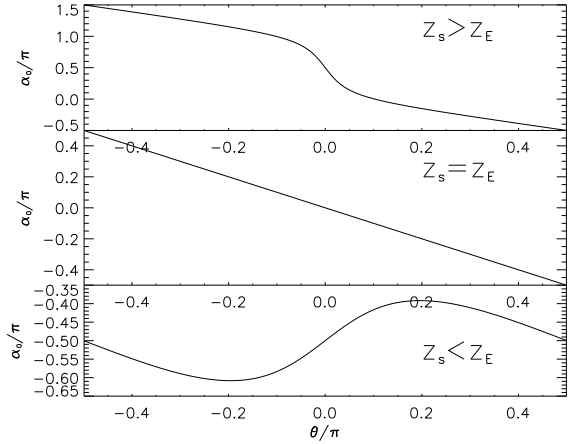}
\caption{ Example of the relationship between $\theta$ and $\alpha_0$ for stationary monochromatic source for  $\xi=1.8$ {\em (top)}, $\xi=2${ \em (middle)}, $\xi=4${ \em (bottom)}.
\label{fig:alphatheta} }
\end{figure}

\appendix
\section{Monochromatic point source: eikonal solution for polytrope model}
\label{sec:appendix}
\subsection{Characteristics} \label{sec:appsolve}
In this section the eikonal equation for polytrope model is solved using the monochromatic point source initial conditions deduced in Section \ref{s:ini_mono_spherical}. For given frequency, $\omega$, and source depth, $z_s,$ we reconstruct the graph of the phase function, $\varphi$, throughout the phase space domain in terms of ray-coordinates using group travel-time as the parameter along the ray. The Jacobian of the transform from ray to Cartesian coordinates is evaluated, and the phase function is reconstructed.

Without  loss of generality,  take $x_s=0$ and $\omega > 0$. By construction (see Section \ref{s:ini_mono_spherical}) , $k_h \geq 0$ and $\theta \in \left( -\frac \pi 2,  \frac \pi 2 \right).$ Note that by our definition (see Figure \ref{fig:geo_setup}) $\theta$ is zero in the horizontal direction and is increasing clock-wise. 
For fixed $z_s, \omega$, the initial values for wavevector components  are determined in terms of the take-off angle, $\theta$, and given by $k_{h0}=k_s(z_s, \omega)\cos \theta,$ $k_{z0}=k_s(z_s, \omega) \sin \theta$. Using the corresponding upper and lower turning points,  in accordance with (\ref{eq:poly_kz_1}), define 
$$a(\theta \arrowvert \omega, z_s)=\frac 1 2 \frac{\omega^2 m} {k_s^2g} (1+\tan^2\theta)$$ and 
$$b^2(\theta \arrowvert \omega, z_s)=a^2-\frac{m(m+2)}{4k_h^2}.$$ 
Then, in view of (\ref{eq:poly_gensolution}), the solution can be written as
\begin{subequations} \label{eq:ap_solution}
\begin{empheq}[left=\empheqlbrace]{align}
x & =\frac{1}{2}\frac{\omega}{k_h}t_g-b \left( \cos \alpha -\cos \alpha_0\right),  \label{eq:alpha_x}\\
z & =a+b \sin \alpha, \label{eq:alpha_z} \\
t & = t_g+t_0, \label{eq:alpha_tg} \\
\alpha (t_g, \theta \arrowvert \omega, z_s) & =\frac{k_h g}{\omega m} t_g + \alpha_0, 
\label{eq:alpha_alpha} \\
\alpha_0(\theta \arrowvert \omega, z_s) & =\sin^{-1} \left( \frac{z_s-a}{b} \right). \label{eq:alpha_alpha0} 
\end{empheq}
\end{subequations}
From this it is immediately clear that for an individual ray upper turning points correspond to the values $\alpha=-\frac{\pi} 2 + 2n\pi$, while lower turning points are given by  $\alpha=\frac{\pi} 2 + 2n\pi$, where $n=0, 1, 2, \ldots$ The horizontal distance between two successive bounces on the surface, called a skip-distance, is $\Delta(k_h, \omega)=2 \pi a=\pi {\omega^2m}/{k_h^2g}=v_{ph}^2\ {\pi m} / g,$ where $v_{ph}= \omega / {k_h}$ is the horizontal phase-speed. The time, $t_\Delta,$ it takes the ray to travel between the bounces is linked to the distance: $\Delta=  g t_\Delta^2 / {4\pi m}.$

Since $\omega > 0,$ $\alpha$ is increasing when $t_g$ is increasing.  With $\omega$ constant on the phase function graph, and $k_h=k_{h0}=k_s\cos\theta$ constant along the ray, we need to find $k_z$ in ray-coordinates to complete the solution (\ref{eq:alpha_x}-\ref{eq:alpha_tg}) to phase space. 
This is found from the condition that  Hamiltonian (\ref{eq:Hamiltonian}) is zero. Indeed from (\ref{eq:poly_kz_1}) and (\ref{eq:alpha_z}) it follows that $k_z= {k_h b \cos \alpha} / {z},$ with the choice of the square root branch taken care of by 
Some care should be taken in choosing the correct branch of the square root, which is addressed via a choice of the inverse sine branch in the definition of $\alpha_0$ (\ref{eq:alpha_alpha0}) as follows.
For given $z_s, \omega$ and $\theta,$ there will be two solutions for parameter $\alpha_0(\theta)$: $\alpha_0^+ \in \left[ -\frac{\pi}{2}, \frac{\pi}{2} \right]$ and $\alpha_0^-=\pi-\alpha_0^+ \in  \left[ \frac{\pi}{2}, \frac{3\pi}{2} \right].$  The former clearly corresponds to rays going down from the source, $z$ increasing with $t_g$, $k_{z0}>0$, and the latter to rays going up ($z$ decreasing, $k_{z0}<0$). Thus, $\alpha_0^+$ is chosen when $\theta>0,$ and  $\alpha_0^-$ when $\theta<0.$ 
To ensure smooth dependence of $\alpha_0$ on $\theta,$ one has to join the two solutions at $\theta=0$, which corresponds to $k_{z0}=0,$ \ie the ray that has source depth, $z_s,$ as one of its turning points. In view of the Proposition \ref{thm:prop} and Corollary \ref{thm:corr} in Section \ref{app:ray_solve}, where a partition depth is deduced and determined for polytrope model, depending on whether the source is located above or below the partition depths $z_E,$ this turning point will be either upper $\left(\alpha_0=\frac{\pi} 2 + 2n\pi\right)$ or lower $\left(\alpha_0=-\frac{\pi} 2 + 2n\pi\right)$.

To treat this mathematically, let us  define 
$$\xi = \left.\frac{\omega^2}{\omega^2-\omega^2_{ac}}\right\arrowvert_{z=z_s}.$$ 
The condition that waves are not evanescent, $k_s^2>0 \Longleftrightarrow  \omega > \omega_{ac}(z_s),$ implies that $\xi \in (1, \infty)$. When $z_s=z_E,$ $\xi=2.$ It is  helpful to rewrite 
\begin{subequations}
\begin{align}
a & =\frac   \xi {2 \cos^2\theta} z_s,  \nonumber \\
b^2 & =\frac{z_s^2}{4\cos^4\theta}\left( (\xi-2)^2+4(\xi-1)\sin^2\theta\right). \nonumber
\end{align}
\end{subequations}
Note that $b = 0$ only when both $\xi=2$ and $\theta=0$. When $b \neq 0$, $b\sin\alpha_0=z_s-a$ and $b\cos\alpha_0=z_s {k_{z0}}/{k_h}=z_s \tan\theta,$ so that
$$
\tan \alpha_0= \frac{1-\frac \xi 2}{\tan \theta}-\frac \xi 2 \tan \theta=  \cot \theta - \frac \xi {\sin 2 \theta}.
$$
Consider the behaviour at $\theta=0,$ as well as $\theta$ approaching edge points, $-\frac \pi 2, \frac \pi 2,$ that can be potentially problematic. There are three cases:
\begin{itemize}

\item $z_s>z_E, \xi \in (1, 2) \Longrightarrow \theta=0$ corresponds to lower turning point. Thus chose $\alpha_0(\theta=0)=\frac \pi 2$. Then, $\alpha_0 \rightarrow {-\frac \pi 2}^-$ as $\theta \rightarrow {\frac \pi 2}^-$; and $\alpha_0 \rightarrow {\frac {3\pi} 2}^+$ as $\theta \rightarrow {-\frac \pi 2}^+$;

\item $z_s<z_E, \xi \in (2, \infty) \Longrightarrow \theta=0$ corresponds to upper turning point. Thus chose $\alpha_0(\theta=0)=-\frac \pi 2$. Then, $\alpha_0 \rightarrow {-\frac \pi 2}^-$ as $\theta \rightarrow {\frac \pi 2}^+$; and $\alpha_0 \rightarrow -{\frac {\pi} 2}^-$ as $\theta \rightarrow {-\frac \pi 2}^+$;
 
\item $z_s=z_E, \xi=2, \tan \alpha_0= - \tan \theta,$ when $\theta \neq 0.$ For $\theta=0$ the value of $\alpha_0$ is not defined, but as $a=z_E,$ let us choose $\alpha_0(\theta=0)=0$ with  $\alpha_0 \rightarrow {-\frac \pi 2}^+$ as $\theta \rightarrow {\frac \pi 2}^-$; and $\alpha_0 \rightarrow {\frac {\pi} 2}^+$ as $\theta \rightarrow {-\frac \pi 2}^+$;
\end{itemize}
Therefore, using formulas for $b, \tan\alpha_0$ and $\cos\alpha_0$, the derivative of $\alpha_0$ by $\theta$ can be expressed in terms of $\xi$:
\begin{empheq}{align}
\dd{\alpha_0}\theta & = \frac{2(\xi-2)-4(\xi-1)\sin^2\theta}{(\xi-2)^2+4(\xi-1)\sin^2\theta} \nonumber \\
& =
-1+\frac{\xi(\xi-2)}{(\xi-2)^2+4(\xi-1)\sin^2\theta}. \label{eq:alpha_0_der}
\end{empheq}
In the case when $z_s=z_E, \xi=2$, the above formula takes form $\dd{\alpha_0}\theta= -1.$ For fixed frequency and source depth, taking into account the information above, relationship (\ref{eq:alpha_0_der}) can be integrated to find
\be
\alpha_0(\theta) =
	\begin{cases}
	\frac \pi 2-\theta+\tan^{-1}\left( \frac \xi{\xi-2}\tan\theta \right)  & z_s> z_E, \xi \in (1, 2),  
	\\
	-\frac \pi 2-\theta+\tan^{-1}\left( \frac \xi{\xi-2}\tan\theta \right)  & z_s< z_E, \xi \in (2, \infty)
	\\
	-\theta & z_s=z_E, \xi = 2. 
	\end{cases}
	\label{eq:alpha_0_sol}
\ee
Representative plots for $\alpha_0$ as function of $\theta$ are shown in Figure \ref{fig:alphatheta}.

\subsection{Caustics} \label{app:jacobian}
Equations (\ref{eq:alpha_x}-\ref{eq:alpha_alpha}) together with with 
\begin{subequations}\label{eq:sol_phase_var}
\begin{empheq}[left=\empheqlbrace]{align}
k_h & =k_s(\omega, z_s)\cos\theta, \label{eq:sol_phase_var1} \\
k_z & =k_h\frac{b \cos\alpha}z,  \label{eq:sol_phase_var2}\\
 \omega & =\const, \label{eq:sol_phase_var3}
\end{empheq}   
\end{subequations}
define the three-dimensional graph of $\varphi$ lying in $\Omega M$ in terms of ray coordinates $(\theta, t_0, t_g).$ 
As $t=t_g+t_0$ and since neither $x$ nor $z$ depend explicitly on $t_0,$ the Jacobian of the coordinate transform can be written as 
\be
\mathcal D=\frac{\partial (x, z, t)}{\partial (\theta, t_0, t_g)}=\left| 
	\begin{array}{ccc}
	\dd{x}{\theta} & 0 & \frac{c^2}{\omega}k_h \\
	 \dd z {\theta} & 0 & \frac{c^2}{\omega}k_z  \\
	0 & 1 & 1\\
	\end{array}
\right| = - \dd{(x, z)}{(\theta, t_g)}
\label{eq:point_jacobian}
\ee
Note that $\mathcal D=0$ when the wave-vector $\left(\ddil x {t_g}, \ddil z {t_g}\right)$ is tangent to the wavefront, \ie collinear to $\left( \ddil x \theta, \ddil z \theta\right).$

When $\theta \in \left( -\frac \pi 2, \frac \pi 2 \right) \Rightarrow k_h \ne 0$, $\alpha$ can be used instead of $t_g$ as a parameter along the ray:
\be
\mathcal D=\frac{\partial (x, z, t)}{\partial (\theta, t_0, \alpha)} \times \frac{\partial (\theta, t_0, \alpha)}{\partial (\theta, t_0, t_g)}
= - \frac {k_h g}{\omega m}\frac{\partial (x, z)}{\partial (\theta, \alpha)}. \label{eq:jacobian_vars}
\ee

Let us consider the general case $z \neq z_E$. Making use of the formulas
\begin{subequations}
\begin{empheq}{align}
\dd{k_h}\theta & =-k_h\tan \theta, \ \ 
\dd a \theta = 2 a \tan \theta, \nonumber \\
\dd b \theta & = \frac{\tan \theta} b (a^2+b^2) {\rm \ when \ }z_s \neq z_E, \nonumber \\
\dd \alpha \theta & = -(\alpha-\alpha_0) \tan \theta + \dd{\alpha_0}\theta, \nonumber
\end{empheq}
\end{subequations}
the partial derivatives with respect to take-off angle are evaluated:
\begin{subequations}\label{eq:part_deriv_caus}
\begin{empheq}[left=\empheqlbrace]{align} 
\dd x \theta  = & (\alpha-\alpha_0) \tan \theta (a - b \sin\alpha) \nonumber \\
& + z_s(1+\tan^2\theta) -\frac{\tan\theta}b(a^2+b^2)\cos\alpha
\nonumber \\
 & +b\sin\alpha\dd{\alpha_0}\theta 
\label{eq:part_deriv_caus1} \\
\dd z \theta  = &  2a\tan\theta+\frac{\tan\theta}b(a^2+b^2)\sin\alpha  \nonumber \\
 & -(\alpha-\alpha_0) b \cos\alpha \tan\theta + b \cos\alpha \dd{ \alpha_0}\theta 
 \label{eq:part_deriv_caus2} \\
\dd x {t_g}  = & \frac{k_h g}{\omega m} (a + b \sin\alpha), \label{eq:part_deriv_caus3} \\
\dd z {t_g} = & \frac{k_h g}{\omega m} b \cos\alpha \label{eq:part_deriv_caus4}
\end{empheq}
\end{subequations}
Then using $\alpha$ as a parameter along the ray define the scaled jacobian

\begin{align}
J (\theta, \alpha)  & =   \frac{\omega m}{k_h g} \mathcal D =   b \cos\alpha \times \nonumber \\ 
& \times \left( 2a(\alpha-\alpha_0)\tan\theta -a\dd{\alpha_0}\theta+ 	z_s(1+\tan^2\theta)\right)  \nonumber \\
& -\frac{\tan\theta}b\left( (a^3+3b^2a)\sin\alpha+3a^2b + b^3 \right). \label{eq:jacobiana}
\end{align}


\subsection{Reconstructing phase function $\varphi$}

Using $\alpha$ as a parameter along the ray and taking into account (\ref{eq:Hamiltonian}), phase function derivative along the ray (\ref{eq:phase_d}) takes the form
\be
\dder \varphi \alpha = k_h \dder x \alpha + k_z \dder z \alpha - \omega \dder t \alpha = - \frac z {k_h} \frac{\omega_{ac}^2}{c^2}. \nonumber
\ee
This is then integrated to reconstruct the phase function along the ray
\begin{align}
	\varphi  & = \varphi_0-\frac{m(m+2)}{4 k_h} 
	\int_{\alpha_0}^{\alpha} \frac{d\alpha'}
	{a+b \sin \alpha'} \nonumber \\
	& = \varphi_0 - \frac 1 {\sqrt{m(m+2)}}  
	 \left[ \tan^{-1}\left( \frac{a}{\sqrt{z_uz_l}} \tan \frac \alpha 2 + \frac{b}{\sqrt{z_uz_l}}  \right) 
	 \right.
	\nonumber \\	
	& \ \ \ \ \ \ \ \ \ \ \ \ \ \ 
		\left.
	- \tan^{-1}\left( \frac{a}{\sqrt{z_uz_l}} \tan \frac {\alpha_0} 2 + \frac{b}{\sqrt{z_uz_l}}  \right)
	\right]
\end{align}

%% file: rays_mnras_review2_1.bbl
\begin{thebibliography}{45}
\expandafter\ifx\csname natexlab\endcsname\relax\def\natexlab#1{#1}\fi

\bibitem[{{Arnold}(1978)}]{ArnoldBook}
{Arnold} V.~I., 1978, {Mathematical methods of classical mechanics}, {Arnold,
  V.~I.}, ed.

\bibitem[{{Barnes} \& {Cally}(2001)}]{BC2001}
{Barnes} G., {Cally} P.~S., 2001, \pasa, 18, 243

\bibitem[{Bogdan(1997)}]{Bogdan97}
Bogdan T.~J., 1997, Astrophysical Journal, 477, 475

\bibitem[{{Bogdan} \& {Cally}(1995)}]{BogdanCally1995}
{Bogdan} T.~J., {Cally} P.~S., 1995, \apj, 453, 919

\bibitem[{{Chapman} {et~al}\mbox{.}(1999){Chapman}, {Lawry}, {Ockendon}, \&
  {Tew}}]{Chapman99}
{Chapman} S.~J., {Lawry} J.~M.~H., {Ockendon} J.~R., {Tew} R.~H., 1999, SIAM
  Review, 41, 417

\bibitem[{Christensen-Dalsgaard(2003)}]{CDLN}
Christensen-Dalsgaard J., 2003, 268

\bibitem[{{Christensen-Dalsgaard} {et~al}\mbox{.}(1996){Christensen-Dalsgaard},
  {Dappen}, {Ajukov}, {Anderson}, {Antia}, {Basu}, {Baturin}, {Berthomieu},
  {Chaboyer}, {Chitre}, {Cox}, {Demarque}, {Donatowicz}, {Dziembowski},
  {Gabriel}, {Gough}, {Guenther}, {Guzik}, {Harvey}, {Hill}, {Houdek},
  {Iglesias}, {Kosovichev}, {Leibacher}, {Morel}, {Proffitt}, {Provost},
  {Reiter}, {Rhodes}, {Rogers}, {Roxburgh}, {Thompson}, \& {Ulrich}}]{CDModel}
{Christensen-Dalsgaard} J. {et~al.}, 1996, Science, 272, 1286

\bibitem[{{Couvidat} {et~al}\mbox{.}(2006){Couvidat}, {Birch}, \&
  {Kosovichev}}]{Couvidat2006}
{Couvidat} S., {Birch} A.~C., {Kosovichev} A.~G., 2006, \apj, 640, 516

\bibitem[{{Couvidat} {et~al}\mbox{.}(2004){Couvidat}, {Birch}, {Kosovichev}, \&
  {Zhao}}]{Couvidat2004}
{Couvidat} S., {Birch} A.~C., {Kosovichev} A.~G., {Zhao} J., 2004, \apj, 607,
  554

\bibitem[{{Donea}(2011)}]{donea11}
{Donea} A., 2011, \ssr, 158, 451

\bibitem[{{Donea} {et~al}\mbox{.}(1999){Donea}, {Braun}, \&
  {Lindsey}}]{Donea1999}
{Donea} A., {Braun} D.~C., {Lindsey} C., 1999, \apjl, 513, L143

\bibitem[{{Donea} \& {Lindsey}(2005)}]{DL2005}
{Donea} A., {Lindsey} C., 2005, \apj, 630, 1168

\bibitem[{D'Silva(1996)}]{DSilva1996}
D'Silva S., 1996, Astrophysical Journal v.462, 462, 519

\bibitem[{D'Silva \& Duvall(1995)}]{DSilva1995}
D'Silva S., Duvall T.~L., 1995, Astrophys J, 438, 454

\bibitem[{{Duvall} {et~al}\mbox{.}(1993){Duvall}, {Jefferies}, {Harvey}, \&
  {Pomerantz}}]{Duvall93}
{Duvall}, Jr. T.~L., {Jefferies} S.~M., {Harvey} J.~W., {Pomerantz} M.~A.,
  1993, \nat, 362, 430

\bibitem[{{Felsen} \& {Marcuvitz}(1991)}]{FM}
{Felsen} L.~B., {Marcuvitz} N., 1991, {Radiation and Scattering of Waves}. New
  York, USA: Wiley-IEEE

\bibitem[{{Giles}(2000)}]{Giles2000}
{Giles} P.~M., 2000, PhD thesis, STANFORD UNIVERSITY

\bibitem[{{Gough}(1993)}]{gough93}
{Gough} D.~O., 1993, in Astrophysical Fluid Dynamics - Les Houches 1987,
  {J.-P.~Zahn \& J.~Zinn-Justin}, ed., pp. 399--560

\bibitem[{{Gough}(2007)}]{gough2007}
{Gough} D.~O., 2007, Astronomische Nachrichten, 328, 273

\bibitem[{Guillemin \& Sternberg(1990)}]{GeoAsymptotics}
Guillemin V., Sternberg S., 1990, 480

\bibitem[{{Kitiashvili} {et~al}\mbox{.}(2011){Kitiashvili}, {Kosovichev},
  {Mansour}, \& {Wray}}]{Kitiashvili2011}
{Kitiashvili} I.~N., {Kosovichev} A.~G., {Mansour} N.~N., {Wray} A.~A., 2011,
  \apjl, 727, L50

\bibitem[{{Kosovichev}(2006)}]{K2006}
{Kosovichev} A.~G., 2006, \solphys, 238, 1

\bibitem[{{Kosovichev}(2007)}]{Kosovichev2007}
{Kosovichev} A.~G., 2007, \apjl, 670, L65

\bibitem[{{Kosovichev}(2011)}]{Kosovichev2011}
{Kosovichev} A.~G., 2011, \apjl, 734, L15+

\bibitem[{{Kosovichev} \& {Duvall}(1997)}]{KosDuval1997}
{Kosovichev} A.~G., {Duvall}, Jr. T.~L., 1997, in Astrophysics and Space
  Science Library, Vol. 225, SCORe'96 : Solar Convection and Oscillations and
  their Relationship, {F.~P.~Pijpers, J.~Christensen-Dalsgaard, \&
  C.~S.~Rosenthal}, ed., pp. 241--260

\bibitem[{{Kosovichev} \& {Zharkova}(1998)}]{kz1998}
{Kosovichev} A.~G., {Zharkova} V.~V., 1998, \nat, 393, 317

\bibitem[{Kravtsov \& Orlov(1990)}]{KravtsovOrlov}
Kravtsov Y.~A., Orlov Y.~I., 1990, Geometrical Optics of Inhomogeneous Media.
  Berlin, GR: Springer-Verlag

\bibitem[{{Kravtsov} \& {Orlov}(1993)}]{KravstovOrlovCaustics}
{Kravtsov} Y.~A., {Orlov} Y.~I., 1993, {Caustics, Catastrophes and Wave
  Fields}. Berlin, GR: Springer-Verlag

\bibitem[{Lindsey {et~al}\mbox{.}(2011)Lindsey, Braun, Hern{\'a}ndez, \&
  Donea}]{LBH2011}
Lindsey C., Braun D., Hern{\'a}ndez I., Donea A., 2011, in 'Holography.
  Different fields of Application', {F.~A.~M.~Ramirez}, ed., p.~81

\bibitem[{{Lindsey} \& {Braun}(2000)}]{LB2000}
{Lindsey} C., {Braun} D.~C., 2000, \solphys, 192, 261

\bibitem[{Lindsey \& Braun(2000)}]{LDfarside}
Lindsey C., Braun D.~C., 2000, Science, 287, 1799

\bibitem[{Lindsey \& Braun(2004)}]{LB2004}
Lindsey C., Braun D.~C., 2004, The Astrophysical Journal Supplement Series,
  155, 209

\bibitem[{{Lindsey} \& {Donea}(2008)}]{LD2008}
{Lindsey} C., {Donea} A., 2008, \solphys, 251, 627

\bibitem[{{Matthews} {et~al}\mbox{.}(2011){Matthews}, {Zharkov}, \&
  {Zharkova}}]{MZZ2011}
{Matthews} S.~A., {Zharkov} S., {Zharkova} V.~V., 2011, \apj, 739, 71

\bibitem[{{Moll} {et~al}\mbox{.}(2011){Moll}, {Cameron}, \&
  {Sch{\"u}ssler}}]{Moll2011}
{Moll} R., {Cameron} R.~H., {Sch{\"u}ssler} M., 2011, \aap, 533, A126

\bibitem[{{Moradi} {et~al}\mbox{.}(2010){Moradi}, {Baldner}, {Birch}, {Braun},
  {Cameron}, {Duvall}, {Gizon}, {Haber}, {Hanasoge}, {Hindman}, {Jackiewicz},
  {Khomenko}, {Komm}, {Rajaguru}, {Rempel}, {Roth}, {Schlichenmaier},
  {Schunker}, {Spruit}, {Strassmeier}, {Thompson}, \& {Zharkov}}]{Moradi2010}
{Moradi} H. {et~al.}, 2010, \solphys, 267, 1

\bibitem[{{Moradi} {et~al}\mbox{.}(2007){Moradi}, {Donea}, {Lindsey},
  {Besliu-Ionescu}, \& {Cally}}]{Moradi2007}
{Moradi} H., {Donea} A., {Lindsey} C., {Besliu-Ionescu} D., {Cally} P.~S.,
  2007, \mnras, 374, 1155

\bibitem[{P{\'e}rez~Hern{\'a}ndez \& Gonzalez-Hernandez(2010)}]{PH2010}
P{\'e}rez~Hern{\'a}ndez F., Gonzalez-Hernandez I., 2010, The Astrophysical
  Journal, 711, 853

\bibitem[{Schunker \& Cally(2006)}]{SchunkerCally}
Schunker H., Cally P.~S., 2006, Mon Not R Astron Soc, 372, 551

\bibitem[{{Shelyag} {et~al}\mbox{.}(2009){Shelyag}, {Zharkov}, {Fedun},
  {Erd{\'e}lyi}, \& {Thompson}}]{shelyag2009}
{Shelyag} S., {Zharkov} S., {Fedun} V., {Erd{\'e}lyi} R., {Thompson} M.~J.,
  2009, \aap, 501, 735

\bibitem[{{Thompson} \& {Zharkov}(2008)}]{ThompsonZharkov2008}
{Thompson} M.~J., {Zharkov} S., 2008, \solphys, 251, 225

\bibitem[{{Zharkov} {et~al}\mbox{.}(2011{\natexlab{a}}){Zharkov}, {Green},
  {Matthews}, \& {Zharkova}}]{ZGMZ2011}
{Zharkov} S., {Green} L.~M., {Matthews} S.~A., {Zharkova} V.~V.,
  2011{\natexlab{a}}, \apjl, 741, L35

\bibitem[{{Zharkov} {et~al}\mbox{.}(2012){Zharkov}, {Green}, {Matthews}, \&
  {Zharkova}}]{ZGMZ2012}
{Zharkov} S., {Green} L.~M., {Matthews} S.~A., {Zharkova} V.~V., 2012,
  \solphys, 292

\bibitem[{{Zharkov} {et~al}\mbox{.}(2011{\natexlab{b}}){Zharkov}, {Zharkova},
  \& {Matthews}}]{ZH2011}
{Zharkov} S., {Zharkova} V.~V., {Matthews} S.~A., 2011{\natexlab{b}}, \apj,
  739, 70

\bibitem[{{Zharkova} \& {Zharkov}(2007)}]{ZZ2007}
{Zharkova} V.~V., {Zharkov} S.~I., 2007, \apj, 664, 573

\end{thebibliography}
